\documentstyle[12pt,axodraw]{article}
 \def\be{\begin{eqnarray}}
 
 \def\ee{\end{eqnarray}}
\input psfig.sty
\begin{document} 
\pagestyle{empty}
\Huge{\noindent{Istituto\\Nazionale\\Fisica\\Nucleare}}

\vspace{-3.9cm}

\Large{\rightline{Sezione di ROMA}}
\normalsize{}
\rightline{Piazzale Aldo  Moro, 2}
\rightline{I-00185 Roma, Italy}

\vspace{0.65cm}

\rightline{INFN-1335/02}
\rightline{May 2002}

\vspace{1.cm}

\begin{center}{\large\bf Pair term in the Electromagnetic Current within the Front-Form Dynamics:
Spin-0 Case}  
\end{center}

\begin{center} {J. P. B. C. de Melo$^a$, T. Frederico$^b$, E. Pace$^c$ and 
G. Salm\`e$^d$}
\end{center}

\noindent { $^a$ \it Instituto de F\'\i sica Te\'orica, Universidade Estadual Paulista 
0145-900, S\~ao Paulo, SP, Brazil}

\noindent { $^b$ \it Dep. de F\'\i sica, Instituto Tecnol\'ogico da Aeron\'autica, 
Centro T\'ecnico Aeroespacial, 12.228-900, S\~ao Jos\'e dos Campos,
S\~ao Paulo, Brazil}

\noindent {$^c$ \it Dipartimento di Fisica, Universit\`a di Roma "Tor Vergata" and Istituto
Nazionale di Fisica Nucleare, Sezione Tor Vergata, Via della Ricerca 
Scientifica 1, I-00133   Roma, Italy }

\noindent { $^d$ \it Istituto
Nazionale di Fisica Nucleare, Sezione Roma I, P.le A. Moro 2,
 I-00185   Roma, Italy }

\begin{abstract}
The frame and scale dependence of the pair-term  contribution 
to the electromagnetic form factor of a spin-zero  composite 
system  of two-fermions is studied within the Light Front. The form factor 
is evaluated from the plus-component of the current 
in the Breit frame, using for the first time a nonconstant, symmetric ansatz 
for the Bethe-Salpeter
amplitude. The frame dependence
is analyzed by allowing a nonvanishing plus component of the momentum transfer,
while the dynamical scale is set by the masses of the constituents and by 
 mass  and  size of the composite system. 
A transverse momentum distribution, associated with the  Bethe-Salpeter
amplitude, is introduced which allows to  define
strongly and weakly relativistic systems.
In particular, for strongly relativistic systems, the pair term 
vanishes for the
Drell-Yan condition, while is dominant for momentum transfer along 
the light-front direction. For a weakly relativistic system, 
fitted to the deuteron scale, 
 the pair term is negligible up to momentum transfers of 1(GeV/c)$^2$. 
A comparison with results obtained within  the Front-Form Hamiltonian dynamics 
with a fixed number of constituents is also presented.
\end{abstract}

\vspace{2.8cm}
\hrule width5cm
\vspace{.2cm}
\noindent{\normalsize{To appear in {\bf Nucl. Phys. A} }}

\newpage
\pagestyle{plain}

\section{Introduction}

Present knowledge of the structure of hadrons and nuclei  mainly
comes from electroweak form factors, in elastic and transition regimes,  and 
from deep-inelastic structure functions. To perform a meaningful
comparison between theoretical models and the experimental data, one needs 
a description of  the  bound system of interacting
constituents and  a consistent  current operator. 
Within a Hamiltonian approach, the state of the system   is defined 
on a specified hypersurface of the space-time that does not contain timelike
directions. Dirac identified
three space-time hypersurfaces, adequate to define the  state of a
relativistic system,  which  
correspond  to different forms of relativistic
Hamiltonian  Dynamics, namely the  Instant Form,  the Front Form
and the Point Form \cite{dirac}.

In the Front-Form dynamics, the consistency
between the current operator and the state of the hadron system 
has been discussed 
both from a field theoretical point of view and  within approaches
with a fixed number of particles. In  field theory, the state  has an 
infinite number of components in the Fock space\cite{brodsky}. 
However, for practical applications only the lowest Fock component, 
or valence component, is usually modeled and used in the calculations of
electroweak 
form factors. In principle, the infinite set of coupled
eigenvalue equations for the Hamiltonian operator in the Fock space can 
be replaced by  an effective squared mass operator  
acting in the valence sector; 
at the same time, it is possible to express systematically the higher 
Fock-state components of the wave function as functionals of the lower ones
\cite{brodsky,pauli3,pauli4}. The effective electroweak 
current operator to be used with the valence component of the state can be  
consistently derived within the  framework of the
Bethe-Salpeter equation projected at equal light-front time, as recently 
shown  in Refs.\cite{sales1,sales2}. 
(For recentinvestigation on other aspects of the Bethe-Salpeter 
equation within the light front, see, e.g. \cite{miller,jaume}).

If a fixed number of interacting particles is assumed, then a  Front-Form Hamiltonian 
dynamics (FFHD) can be developed.
One can use  the 
Bakamjian-Thomas construction\cite{bt}, where an effective 
interaction in the mass operator can be chosen such that
a unitary representation of the Poincar\'e group is possible. 
Taking the Front-Form spin operator as the 
free one and an interaction in the mass operator that commutes with such a
spin operator, one is able to explicitly construct generators of the 
Poincar\'e group\cite{kp,FLev}. In this case, the eigenfunctions
of the mass operator are normalized to one, differently from
the field theoretical valence component, which has in general a probability
less than one\cite{brodsky}. A wide number of papers have been devoted to the
evaluation of properties of hadrons and nuclei  within 
the  Front-Form Hamiltonian dynamics with a finite number of particles 
(see, e.g. Refs.
\cite{coester,chung,we,CP,salme,card,jaus1}, 
just to give a short account of some previous works). In particular,
 it should be noted that FFHD for the two-nucleon case yields the possibility 
 to retain the
large amount of successful phenomenology 
developed within a nonrelativistic
approximation (see, e.g., \cite{coester,levprl}). Usually, to obtain
  electroweak  form factors
of nuclear systems or hadrons,
 the Drell-Yan condition on the momentum transfer, i.e.  $q^+=q^0+q^3=0$, 
 and the matrix elements of the 
 plus component of the
current, $j^+$,
are adopted. The physical argument often advocated in favour of these
assumptions is that the production of pairs from the incoming photon 
(nonvalence or pair-term contribution) 
is suppressed in the Drell-Yan frame by
light-front momentum conservation (see, e.g., \cite{FS}). 
This was indeed proved in 
schematic covariant field theoretical models, for spin-zero  two-boson 
composite systems\cite{sawicki} and for the pion with pseudo-scalar 
coupling to the quarks\cite{pipach}. However, for
spin-one systems the pair term   survives and 
contributes to $j^+$ in the Drell-Yan frame \cite{pach98,pach99,naus98}. 
Furthermore, in this case
the pair term is necessary to keep the rotational invariance of the form 
factors
for spin-one particles. Within FFHD,
in the Drell-Yan frame, the matrix elements 
of the plus component of the current should fulfill the so called
angular condition \cite{inna}. This constraint is not satisfied by a calculation
considering only the Front-Form wave function and the impulse approximation. 
This approximation implies an ambiguity in extracting form factors for spin-one systems, 
as shown for the deuteron\cite{inna,coester,to91} and for 
the rho-meson \cite{keisterrho,inna95,pach97}. 
To solve this problem,  related to the lacking of covariance
 for  the impulse approximation current operator, some physically inspired
combinations of matrix elements were proposed to extract form factors 
from the current, as the "good-component" approach\cite{to93}
or the elimination of spurious contributions to the form 
factors\cite{karmanov}.

To avoid the difficulties associated with these ambiguities, another
approach was proposed in Ref.\cite{lev98} and applied to the 
deuteron\cite{levprl,lev00,pace01}. 
In these works, it was shown  that 
a current operator which satisfies the requirements of Poincar\'e, parity and 
time-reversal covariance,
as well as hermiticity and current conservation, can be obtained from
a one-body operator in the Breit frame with 
momentum transfer along the z-direction. Then, for any hadron system 
the electromagnetic form factors can be calculated from the wave function 
without ambiguities. 

This major development in the calculation of the form factors
from the Front-Form wave function is, however, confronted with the fact 
that, even in simple field-theoretical models, the pair production mechanism
contributes in the $q^+\ne 0$ Breit frame, as was discussed in 
the calculation of the pion form factor from $j^+$  in Ref. \cite{bakker01}.
 Therefore, within  an approach with a 
fixed number of particles such a mechanism
should be taken into account through an effective two-body current.

The aim of the present work is to  investigate in a covariant model, based on a
nonconstant, symmetric vertex function,  the
effect of the pair term in the evaluation of electromagnetic 
form factors for a pion-like system composed by  two
identical fermions. It is well
known that, to avoid divergences in the evaluation of the covariant 
triangle-diagram for
the form factor some regularization has to be introduced. In Refs.
\cite{pipach,pach98,pach99} a pion-$q \bar{q}$
vertex function, non symmetric in the four-momenta of the quarks, was adopted.
However, a non symmetric vertex cannot be considered a realistic approximation
of a $q \bar{q}$ bound-state amplitude and phenomenological problems arise 
(e. g., the form factor and the  weak decay constant cannot be simultaneously
reproduced) \cite{pipach}. Another approach has been proposed in  Ref. 
\cite{bakker01}, where the fermion loop was
regulated by considering a non-local photon vertex.
In the present work,  for describing the momentum part of 
the coupling between the constituents and the spin-0 system, we use 
a covariant model with the following form of the 
vertex function 
\begin{equation}
\Lambda(k,P)=
\frac{C}{(k^2-m^2_{R} + \imath \epsilon)}+
\frac{C}{((P-k)^2-m^2_{R}+
\imath \epsilon)} \ .
\label{vertex}
\end{equation}
Differently from \cite{pipach,pach98,pach99}, we adopt a vertex function
which is symmetric by the exchange of the  momentum of the two
fermions and  implies a light-front valence wave function with the
same property, as  shown in the following Sections. In this way,
we are simulating the symmetry properties of a Bethe-Salpeter
amplitude derived from quantum field theory. 
The other main ingredient of our covariant model,
for calculating the form factor of the composite system, is the electromagnetic current, that
is taken in impulse approximation.
It should be pointed out that within
our approach such a current 
is conserved (see Sect. II).
In the spirit of Ref. \cite{bakker01},
we study the 
importance of the pair diagram evaluated in different Breit frames, which
differ for the direction of the spatial part of the momentum transfer. 
 We analyze and compare  two  systems that have relativistic or nonrelativistic nature,
respectively. In order to better define this feature, 
we first construct the valence wave function from the symmetric ansatz for
the Bethe-Salpeter
amplitude and  then, from this
wave function,
we build up the transverse-momentum distribution of the constituents.
This momentum distribution plays a twofold role, allowing one: i) to
make contact betwen our covariant model and dynamical models 
of the composite system, developed within
approaches with a
fixed number of particles, and,
more important, ii) to quantitatively define the two limiting cases
that we will consider, namely the strongly and weakly relativistic systems. 
In the first case, we make calculations for a 
spin-0 model, well suited for the pion. Then
we compare
these results with the ones obtained with a realistic pion wave function
generated by a potential able to describe the meson spectroscopy\cite{gi}. 
In the second case, we artificially adjust the transverse momentum 
distribution of the model to the
deuteron scale, to get 
insight into the pair-term contribution
for a weakly relativistic system with a small  average transverse momentum for the 
constituents (getting rid of the lengthy algebra of the 
$m_1 \ne m_2$ case, corresponding to the actual case of charged heavy mesons). 

The paper is organized as follows. In Sec. II, our  model for
the spin-zero, two-fermion system with a symmetric vertex function is presented. 
In Sec. III,  the light-front valence wave function for the 
covariant model is introduced, as well as the corresponding
transverse momentum distribution and the corresponding elastic form factor. 
In Sec. IV, numerical results are presented for i)  the electromagnetic form factor of our model,
ii)  the separate
contribution of the pair term and iii)  the  form factor corresponding to the valence
component. 
In Sec. V, we draw our conclusions.

\section{Electromagnetic form factor of a pion-like system}

In our model, the electromagnetic current for  a two-fermion composite 
system with spin equal to 0
- i.e., a pion-like system considered as $q\bar{ q}$ bound state - 
is calculated in one-loop approximation (triangle diagram), modelling the
Bethe-Salpeter amplitude through a symmetric vertex function in
momentum space with a pseudoscalar coupling between pion 
and quark degrees of freedom. This coupling is suggested 
by a simple effective Lagrangian (see, e.g. \cite{tob92})
\begin{equation}
{{\cal L}_I= - \imath g \vec\Phi \cdot \overline q \gamma^5 \vec 
\tau q  \, }.  \label{lain}
\end{equation}
 The coupling constant $g$ is   given by the
Goldberg-Treiman relation at the quark level, namely $g=m/f_\pi$, with
$m$  the mass of the constituents and $f_\pi$ the pion decay
constant.

The electromagnetic current of  $\pi^+$ is obtained from the covariant 
expression corresponding to the triangle diagram (see, e.g., \cite{Bro69} and
\cite{Riska94}): 
\begin{eqnarray}
j^\mu &=&-\imath 2 e \frac{m^2}{f^2_\pi} N_c\int \frac{d^4k}{(2\pi)^4} Tr[S(k)
\gamma^5 S(k-P^{\prime}) \gamma^\mu S(k-P) \gamma^5 ]\Lambda(k,P^{\prime})
\Lambda(k,P)  \ ,  \label{jmu}
\end{eqnarray}
where $\displaystyle S(p)=\frac{1}{\rlap\slash p-m+\imath \epsilon} \, $, 
 $N_c=3$ is the number of colors, $P^{\mu}$ and 
 $P^{\prime {\mu}}=P^{\mu}+q^{\mu}$ are the initial and
 final momenta of the system,  $q^{\mu}$ is the momentum transfer and $k^{\mu}$
   the spectator quark momentum.  The factor 2 stems from isospin algebra.
(Current conservation can be easily proven in the Breit frame:
after performing the trace in $q\cdot j$,
one notices that the integrand of the 
resulting expression is odd by changing $\vec k \rightarrow - \vec k$; this means that
$q\cdot j$ is zero.)

In Eq. (\ref{jmu}), we introduce the symmetric vertex function of Eq. 
(\ref{vertex}). This vertex function
  produces a light-front wave function symmetric by the interchange of 
quark and antiquark momenta,  and is not affected by
the conceptual difficulties associated with the use of the nonsymmetric 
regulator, as mentioned in the Introduction   (see 
Refs.\cite{pach97,bakker01}). Since we have not specified the dynamics which drives
 the Bethe-Salpeter amplitude,
we have to resort to a physical condition for normalizing the Bethe-Salpeter 
vertex. As a matter of fact, the normalization constant $C$ in the vertex 
function, Eq. (\ref{vertex}), 
is fixed by 
imposing the charge normalization condition (i.e. the pion form factor 
at zero momentum transfer must be
equal to 1).

In our analysis we consider  Breit frames, where the momentum transfer $q^\mu$
 has the spatial component parallel to the
$z-x$ plane. By using  Front-Form variables, i.e. $k^+=k^0+k^3 \ , k^-=k^0-k^3 \ , 
\vec k_\perp\equiv(k^1,k^2)\, $ one has  
\begin{eqnarray} 
 && q^+=-q^-=\sqrt{-q^2}\sin \alpha, \quad
q_x=\sqrt{-q^2}\cos \alpha, \quad  q_y=0 \nonumber \\
&& q^2=q^+q^--(\vec q_\perp)^2
\label{alpha} 
\end{eqnarray}
The Drell-Yan condition $q^+=0$ is recovered with $\alpha=0$,
while the $q^+=\sqrt{-q^2}$ condition \cite{lev98} comes with $\alpha=90^o$. 
(Note that the angle $\theta$ of 
Ref.\cite{bakker01} corresponds to $\alpha+90^o$). The initial  and 
final 
momenta of the composite spin-0 bound state are:  
$P^0=E=E^\prime=\sqrt{m^2_B -q^2/4}$, 
$\vec P^{\prime}_\perp=-\vec P_\perp=\vec q_\perp/2$ and
$P^{\prime}_z=-P_z=q^+/2$.

As well known, the pion form factor can be  extracted from the 
covariant expression: 
\begin{equation}
j^\mu = e (P^{\mu}+P^{\prime \mu}) F_\pi(q^2) \ .
\label{full}
\end{equation}
If covariance and current conservation are fulfilled in a  model calculation, 
 one  can obviously use any frame  and any nonvanishing component of the current
to calculate
the electromagnetic form factor.
We calculate the  form factor  for our pion model defined by 
 Eq. (\ref{vertex}) and  Eq. (\ref{jmu}),
 using the plus component of the
current in the Breit frame with $\vec q$ in the $z-x$ plane. In the evaluation of the
 form factor one can single out two
nonvanishing contributions  
in Eq. (\ref{jmu}) 
\cite{sawicki,pipach,bakker01,ji1}:
\begin{eqnarray}
F_\pi(q^2)=F^{(I)}_\pi(q^2,\alpha)+F^{(II)}_\pi(q^2,\alpha) \ ,
\label{ffactor}
\end{eqnarray}
where $F^{(I)}_\pi(q^2,\alpha)$ has the loop integration on $k^+$  constrained by
$0~\le~ k^+ \ <  \ P^+$, see the light-front time-ordered diagram in Fig. 1(a),
 while $F_\pi^{(II)}(q^2,\alpha)$ has the loop integration 
on $k^+$  in the interval $P^+~\le~ k^+ ~ \le ~ P^{'+}$, see Fig. 1(b). The  valence component of the pion
 contributes to  $F_\pi^{(I)}(q^2,\alpha)$ only,
but in our model it  does not give
the full result in $0~\le~ k^+ \ <  \ P^+$, as  
 discussed in detail 
 in Sec. IV. The
component $F_\pi^{(II)}(q^2,\alpha)$ of the form factor is the contribution 
of the pair production mechanism from an incoming virtual 
photon with $q^+ \ > \ 0$.  

The two contributions to the form factor obtained from $j^+$
are given by the following expressions
\begin{eqnarray}
F_\pi^{(I)}(q^2,\alpha)&=& - \imath
\frac{m^2}{(P^{+}+P^{\prime +})f^2_\pi} 
\frac{N_c}{(2\pi)^4} 
 \int \frac{d^{2} k_{\perp} d k^{+} d k^-\theta (k^+)\theta(P^+-k^+)}
{k^+(P^+-k^+) (P^{^{\prime}+}-k^+)} 
\Pi(k,P,P')
\label{FI}
\end{eqnarray} 
and
\begin{eqnarray}
F_\pi^{(II)}(q^2,\alpha)&=& - \imath
\frac{m^2 }{(P^{+}+P^{\prime +})f^2_\pi} 
\frac{N_c}{(2\pi)^4} 
 \int \frac{d^{2} k_{\perp} d k^{+} d k^-\theta (k^+-P^+)
\theta (P^{\prime +}-k^+)}
{k^+(P^+-k^+) (P^{^{\prime}+}-k^+)} \Pi(k,P,P') 
\label{FII}
\end{eqnarray} 
where
\begin{eqnarray}
\Pi (k,P,P')&=& \frac{Tr[{\cal O}^+]\Lambda(k,P) \Lambda(k,P^\prime)} 
{ (k^- - k^-_{on}+\imath \epsilon)
(P^- - k^- -(P-k)^-_{on}+ \frac{\imath\epsilon}{P^+ - k^+})}
\nonumber \\ &\times &
\frac{1}
{(P^{\prime -} - k^- -(P^\prime-k)^-_{on}+\imath \epsilon)}
\ ,
\label{pi}
\end{eqnarray} 
with  the on-energy-shell values of the individual momenta given by
\begin{eqnarray}
k^-_{on}=\frac{k_{\perp}^{2}+m^2}{k^+} \ , \
(P-k)^-_{on}=\frac{(P-k)_{\perp}^{2}+m^2}{P^+-k^+} ~ ~{\rm{and}} ~ ~
 (P^\prime-k)^-_{on}=\frac{(P^\prime-k)_{\perp}^{2}+m^2}{P^{\prime +}-k^+} \ .
\label{onek}
\end{eqnarray} 
In Eq. (\ref{pi}), the trace $Tr[{\cal O}^+ ]$  of the operator 
\begin{eqnarray}
{\cal O}^+= (\rlap\slash k +m)\gamma^5(\rlap\slash k - \rlap\slash P^\prime +m)
\gamma^+ (\rlap\slash k - \rlap\slash P + m) \gamma^5 \ ,
\label{O}
\end{eqnarray} 
is given by:
\begin{eqnarray}
\frac14 Tr[{\cal O}^+ ]  &=& 
-k^-(P^{\prime +}-k^+)(P^{+}-{k^+})
+(k^2_\perp+m^2)(k^+-P^+-P^{\prime +})
\nonumber \\
&-&\frac12 \vec k_\perp \cdot 
(\vec P^{\prime }_\perp-\vec P_\perp)
(P^{\prime +}-P^+)+\frac14 k^+q^2_\perp \ .
\label{tr}
\end{eqnarray}

The explicit form of the symmetric regulator function in Front-Form
 momentum coordinates is given by
\begin{eqnarray}
\Lambda (k,P)&=& C  \left[k^+(k^--\frac{k^2_\perp+m^2_R-\imath \epsilon}{k^+})
\right]^{-1}
\nonumber \\
&+& 
C \left[(P^+-k^+)(P^--k^--\frac{(P-k)^2_\perp+m^2_R-\imath \epsilon}
{P^+-k^+})\right]^{-1} \ 
\label{llf}
\end{eqnarray}
where the position of the poles for $k^-$ clearly appears.

The detailed forms of $F^{(I)}_\pi$ and $F^{(II)}_\pi$, after integrating 
over $k^-$,  are given in the Appendices A and B,
respectively. In what follows, we will discuss some general features of
Eq. (\ref{FI}) and Eq. (\ref{FII}).
 
Since the integration range of 
$k^+$  is
$0\ \le \ k^+ \ <  \ P^+$ in Eq. (\ref{FI}) and $P^+\ \le \ k^+ \ \le \ P^{'+}$ in
Eq. (\ref{FII}),
then, the sign of the imaginary part 
of some of the poles in the $k^-$-complex plane 
 changes (see  Eq. (\ref{pi}) and Eq. (\ref{llf})). The poles that have their
imaginary part modified are
\begin{eqnarray}
k^-_{(1)}=P^{-} -(P-k)^-_{on}+\frac{\imath \epsilon}
{P^+-k^+}
=P^--\frac{(P-k)_\perp^2+m^2}
{P^+-k^+} +{ \imath \epsilon \over 
{P^+-k^+}}\ ,
\label{k1}
\end{eqnarray}
and  
\begin{eqnarray}
k^-_{(2)}=P^--{(P-k)_\perp^2+m_R^2 \over {P^+-k^+}}+{ \imath \epsilon \over 
{P^+-k^+}} \ .
\label{k2}
\end{eqnarray}
The last one comes from the vertex function, Eq. (\ref{llf}).
The difference in the  sign  of the imaginary parts of $k^-_{(1)}$ and
$k^-_{(2)}$ for the intervals $0\ \le \ k^+ \ <  \ P^+$  and
 $P^+\ \le \ k^+ \ \le \ P^{'+}$ 
is the mathematical signature of the pair production mechanism,
which appears just in the second interval.

The sum of the contributions $F^{(I)}_\pi(q^2,\alpha)$  and 
$F^{(II)}_\pi(q^2,\alpha)$ 
 yields the covariant result, dependent upon $q^2$ only.
Then the different directions of $\vec q$ in the
Breit frame can only change the  values of $F_\pi^{(I)}(q^2,\alpha)$ 
and $F_\pi^{(II)}(q^2,\alpha)$,
but not their sum. For instance, by chosing $q^+=0$ (i.e. $\alpha=0$) 
$F_\pi^{(II)}(q^2,\alpha)$ vanishes and therefore 
$F_\pi^{(I)}(q^2,\alpha)$ alone gives the whole, covariant result
\cite{sawicki}.

It is interesting to note that in our model 
the pair term is linear in $q^+$, for small $q^+$,
as one can verify by direct inspection of the structure of Eq. (\ref{FII}),
once the $k^-$ integration is performed. As a matter of fact,
 the contour for the 
Cauchy integration in the calculation of the pair diagram
can be closed in the upper complex
$k^-$ semi-plane (see Appendix B), and consequently the poles in the integrand of 
Eq. (\ref{FII}) are 
\begin{eqnarray}
k^-_{(3)}=P^{\prime -} -(P^\prime-k)^-_{on} +
\imath \epsilon=
P^{\prime -}- 
\frac{(P^\prime-k)_{\perp}^{2}+m^2}{P^{\prime +}-k^+}+\imath \epsilon \ ,
\label{k3}
\end{eqnarray}
 and
\begin{eqnarray}
k^-_{(4)}=P^{\prime -}-\frac{(P^\prime-k)_{\perp}^{2}+m^2_R
}{P^{\prime +}-k^+} +\imath \epsilon\ .
\label{k4}
\end{eqnarray}
The first pole, Eq. (\ref{k3}), comes from the last factor of 
Eq. (\ref{pi}), while the second pole, Eq. (\ref{k4}) comes 
from $\Lambda(P^{\prime},k)$. The position of both  poles 
in $k^-$ are $\sim (q^+)^{-1}$ in the limit of $q^+\rightarrow 0$.
Then, in order to find the dependence of $F^{(II)}_{\pi}(q^2,\alpha)$ on $q^+$
in this limit, it is enough to count the power of
$q^+$ in  Eq. (\ref{FII}), when   the residues are evaluated. 
The phase-space factor in the denominator
of Eq. (\ref{FII})  is of the order of $(q^+)^2$. Then, let us
consider the trace, Eq. (\ref{tr}), with the proper values of $k^-$ (see
Appendix B). The first term and third one are $\sim q^+$, 
the  second term and the fourth one are of order  $(q^+)^0$. 
Then the trace is
of order $(q^+)^0$.  Evaluating the contribution to the residues from the
remaining part of Eq. (\ref{pi}), we found 
that $\Pi(k,P,P')$ is of the order of $(q^+)^2$. Therefore 
the integrand  goes to a constant for
$q^+\rightarrow 0$, and thus $F^{(II)}_{\pi}(q^2,\alpha)$ is proportional to
$q^+$ in this limit, because of the range of  the $k^+$ integration.

A relevant feature in the analysis of the form factor is  given by the
presence of a contribution  which is instantaneous  in the light-front time, 
 and is produced by the
instantaneous term present in the Dirac propagator. As a matter
of fact,  the Dirac propagator can be decomposed using the Front-Form momenta
as follows \cite{brodsky}
\begin{eqnarray}
\frac{\rlap\slash{k}+m}{k^2-m^2+\imath \epsilon}=
\frac{\rlap\slash{k}_{on}+m}{k^+(k^--k^-_{on}+\frac{\imath \epsilon}{k^+})}
+\frac{\gamma^+}{2k^+} \ ,
\label{inst}
\end{eqnarray}
 where the
second term, proportional to $\gamma^+$, is an
instantaneous term in the light-front time. It should be pointed out that the
instantaneous contribution to the form factor is produced only 
by  the spectator fermion. 
Indeed, the
  instantaneous terms pertaining to
the other propagators do not contribute, because  of the factor 
$\gamma^+$  from the current and the property $(\gamma^{+ })^2=0$.

In our symmetric model, the instantaneous term of Eq. (\ref{inst})
contributes both to $F_\pi^{(I)}(q^2,\alpha)$ and $F_\pi^{(II)}(q^2,\alpha)$,
 due to the analytic structure of the 
symmetric vertex function of Eq. (\ref{vertex}).
These contributions are of
nonvalence nature, since they cannot be  reduced to the impulse approximation
with the valence wave function.

\section{Valence Light-front wave function}

The valence component of the light-front wave function
can be obtained from the Bethe-Salpeter amplitude eliminating the
relative light-front time, i.e.,
constraining to equal light-front time the external 
space-time coordinates of the two fermions,  after dropping {\em the instantaneous terms of
the external Dirac propagators}\cite{sales2}. Physically, 
in the external legs
only light-front propagating particles are allowed.
It is worth noting that the effect of the instantaneous terms, which are 
present in a Bethe-Salpeter 
approach, could be included in  effective (many-body) operators to be used together
with the valence wave function \cite{sales2}. 

In the present model
the Bethe-Salpeter amplitude is
\begin{eqnarray}
\Psi (k,P) = {m \over f_{\pi}}~\frac{\rlap\slash{k}+m}{k^2-m^2+\imath \epsilon}
\gamma^5 \Lambda (k,P) 
\frac{\rlap\slash{k}-\rlap\slash{P}+m}{(k-P)^2-m^2+\imath \epsilon}
\ .
\label{bsa}
\end{eqnarray} 
The momentum part of the valence component of the light-front wave function, 
$\Phi(k^+,\vec k_\perp; P^+,\vec P_\perp)$, can be obtained by eliminating 
out from Eq. (\ref{bsa}):
 i)  the instantaneous terms, ii) 
 the factors containing gamma matrices in the numerator, and iii) the 
 $k^+$ and $(P^+-k^+)$ factors appearing in the denominator. Then, 
after introducing the explicit expression for $\Lambda$, 
one has to integrate over $k^-$, viz.
\begin{eqnarray}
\Phi(k^+,\vec k_\perp; P^+,\vec P_\perp)&=&\imath ~{\cal N}
\int \frac{dk^-}{2\pi}
\frac{1}{(k^--k^-_{on}+ \frac{\imath\epsilon}{k^+})
(P^--k^--(P-k)^-_{on}+\frac{\imath\epsilon}{P^+-k^+})}
\nonumber \\
&\times&
\left( \frac{1}{k^2-m_R^2+\imath \epsilon}
+\frac{1}{(P-k)^2-m_R^2+\imath \epsilon} \right) \ ,
\label{wf1}
\end{eqnarray}
where ${\cal N}$ is a normalization factor
\begin{eqnarray}
 &&{\cal N}=\sqrt{N_c}~C~\frac{m}{f_\pi} \ .
 \label{nwf} 
\end{eqnarray}  
Performing the $k^-$ integration in Eq. (\ref{wf1}), 
one has
\begin{eqnarray}
\Phi(k^+,\vec k_\perp; P^+,\vec P_\perp)=~
\frac{P^+}{m^2_\pi-M^2_{0}}&&\left[\frac{{\cal N}}
{(1-x)(m^2_{\pi}-{\cal M}^2(m^2, m_R^2))} \right.
\nonumber \\
&&\left. +\frac{{\cal N}}
{x(m^2_{\pi}-{\cal M}^2(m^2_R, m^2))} \right]
\ ,
\label{wf2}
\end{eqnarray}
where $x=k^+/P^+$, with $0 \ \le \ x \ \le \ 1$; 
${\cal M}^2(m^2_a, m_b^2)= \frac{k^2_\perp+m_a^2}{x}+\frac{%
(P-k)^2_\perp+m^2_{b}}{1-x}-P^2_\perp \ ;$
and the square of the free mass is $M^2_0 ={\cal M}^2(m^2, m^2)$. 
Since we have
chosen a nonconstant, symmetric $\Lambda$,  
a second term appears in Eq. (\ref{wf2}), differently from Ref. 
\cite{pipach}, and then the 
momentum part of the wave function becomes symmetric by the exchange of 
the momenta of the two constituents. 

By using only the valence component and generalizing the results of Refs.
\cite{pipach,tob92},
the electromagnetic form factor, $ F_\pi^{(WF)}$, evaluated in the
Breit frame is written as follows 
\begin{eqnarray}
F_\pi^{(WF)}(q^2,\alpha)= \frac{1}{2\pi^3(P^{\prime +}+P^+)}
 &&\int \frac{d^{2} k_{\perp} d k^{+} \theta (k^+)\theta(P^+-k^+)}
{k^+(P^+-k^+) (P^{^{\prime}+}-k^+)} 
\Phi(k^+,\vec k_\perp;P^{\prime +},\frac{\vec q_\perp}{2})
\nonumber \\
&\times&
\left [ k^-_{on}P^+P^{\prime +}+\frac12 \vec k_\perp \cdot \vec q_\perp 
(P^{\prime +}-P^{ +})-\frac14 k^+q^2_\perp \right ]
\nonumber \\
&\times&
\Phi(k^+,\vec k_\perp;P^{ +},-\frac{\vec q_\perp}{2})
\ ,
\label{Fwf}
\end{eqnarray} 
with $k^-_{on}=(k^2_\perp+m^2)/k^+$ (see Eq. (\ref{onek})).  
Once the 
normalization constant $C$ is determined from the condition
$F_\pi(0)=1$ in Eq. (\ref{ffactor}), the value of $F_\pi^{(WF)}$ for $q^2=0$ 
yields 
the probability (independent of $\alpha$) of
the valence $q\overline q$ component in the pion, $\eta$.
It should be pointed out that a
value of $\eta~<1$ is expected, if the non-valence contributions
are important, and this is just what occurs in our model (see  Sect. IV for
details), differently from the case of a nonsymmetric ansatz for the
Bethe-Salpeter amplitude, where $\eta~=1$ \cite{pipach}. 

Equation (\ref{Fwf})  represents a point of contact between a field theoretical
approach and  the 
FFHD with a fixed number of particles,  adopting an impulse
approximation current operator. 
Indeed, including a proper factor in $\Phi$, namely $\sqrt{M_0/P^+}$ and 
 normalizing the wave function one can
recover the  FFHD  expression for the form factor, $F^{FFHD}_\pi$. In particular,   
putting  $\alpha=0^o$, one 
obtains the standard FFHD  expression  in the 
frame $q^+=0$ \cite{chung}. 

The valence component of the light-front wave function, Eq. (\ref{wf1}), 
is not an eigenfunction of the
total angular momentum, since it is only one of the components of the pion state 
in the Fock space.  Therefore, we cannot directly compare the valence wave function 
of the present model with the wave functions 
corresponding to approaches with a fixed number of particles \cite{card}, 
which are eigenfunctions of
the intrinsic angular momentum \cite{kp}. However, in order to make contact with
dynamical models,
we introduce the transverse momentum probability density
\begin{eqnarray}
f(k_\perp)= \frac{1}{4\pi^3 m_\pi} \int_0^{2\pi} d\phi 
\int^{m_\pi}_0 \frac{d k^{+}M_0^2}
{k^+(m_\pi-k^+)} \Phi^2(k^+,\vec k_\perp;m_\pi,\vec 0)
\ ,
\label{prob1}
\end{eqnarray} 
By integrating $f(k_\perp)$ over $\vec{k}_\perp$, 
we obtain the probability of the valence component in the pion:
\begin{eqnarray}
\eta=\int^\infty_0 dk_\perp k_\perp f(k_\perp)
\ .
\label{prob2}
\end{eqnarray}
 
Furthermore, the transverse momentum probability density vs 
$k_\perp/m$ results to be very useful   
for a quantitative definition of strongly and weakly relativistic 
composite systems. Indeed high values of the transverse momentum distribution 
for $k_\perp /m > 1$ are a distinctive feature of a strongly relativistic
system.
In particular, in Sect. IV  $f(k_\perp)$ will allow us to investigate 
the influence of the dynamical scale of the system on the role played by the 
 pair term. 

 Another relevant quantity to be used for constraining the parameters of
 our model is
the pion decay constant, $f_\pi$. 
It is defined through the matrix element of the partially
conserved axial-vector current 
\begin{eqnarray}
&& P_\mu <0|A^\mu_i |\pi_j>= \imath~ m_\pi^2 f_\pi \delta_{ij} \ . 
\end{eqnarray}  
Following Ref.\cite{tob92}, we take $A^\mu_i = \bar{q} \gamma^\mu \gamma^5 
\frac{\tau_i}{2} q$ and adopt our ansatz for the  
 pion-$\bar{q} q$ vertex function. In this way we obtain 
\begin{eqnarray}
\imath P^2 f_\pi &=& \frac{m}{f_\pi} N_c\int \frac{d^4k}{(2\pi)^4} Tr[%
\rlap\slash P \gamma^5 S(k) \gamma^5 S(k-P)] \Lambda(k,P)  \ ,  \label{f_pi}
\end{eqnarray}
and integrating on  $k^-$, one gets $f_\pi$ in terms of  
the valence component of our model:
\begin{eqnarray}
f_{\pi} = \frac{m ~\sqrt{N_c}}{4\pi^3}
\int \frac{d^{2} k_{\perp} d k^+ } {k^+(m_\pi-k^+)}  
\Phi(k^+,\vec k_\perp;m_\pi,\vec 0) \ .
\label{fpi}
\end{eqnarray}

\section{Numerical Results}

\subsection{Pion model}

In our model, we have two free parameters for the pion:
 the constituent quark mass, $m$,
and the regulator mass, $m_R$. The constituent quark mass is
chosen as $m=0.220$ GeV,
adequate for the meson phenomenology\cite{gi,tob92,inna94}.
The regulator mass
$m_{R}=0.6$ GeV is found by fitting 
Eq. (\ref{fpi}) to the experimental value $f^{exp}_\pi=92.4$ MeV.
For the pion mass we use the experimental value of 0.140 GeV. 
As a consequence, the charge radius, obtained from
$\langle r^2 \rangle= 6\frac{\partial}{\partial q^2} F_\pi$,
 comes out to be 0.74 fm, which is about $10 \%$ larger than the experimental
 value ($r_{exp}=0.67\pm 0.02$ fm \cite{amen}). 

In Fig. 2, the results for the
pion form factor  are shown and compared to the data of
Refs.\cite{tj,cea,corn1,corn2,bebek}. The full-model 
 calculations, Eq. (\ref{ffactor}),
nicely agree with the new data for the pion form factor \cite{tj}.
Therefore, our model, based on a  non-constant, symmetric vertex can reproduce 
the form factor data 
consistently
with the experimental value of $f_\pi$, while for the
nonsymmetric regulator this was not possible\cite{pipach}. 
Remarkably, in order  to reproduce simultaneously
$f_\pi$ and the experimental form factor  within our model,
the constituent quark mass has to be chosen 
 in the range between 0.2 and 0.3 GeV. Compatibility
between form factor and decay constant has been also achieved in
\cite{bakker01},
where a constant pion-$q\bar{q}$ vertex and a nonlocal photon vertex were
adopted.

The  form factor in the Breit  frame
with $\alpha=0^o$ ($q^+=0$), where the pair-term contribution
is zero, is identical to the sum of $F_\pi^{(I)}$ and
$F_\pi^{(II)}$, calculated in the frame corresponding to  $\alpha=90^o$, 
as it must be
for a covariant model. In Fig. 2,
  $F_\pi^{(I)}$ and $F_\pi^{(II)}$, for  $\alpha=90^o$, 
 are also shown. Differently 
from the case $\alpha=0^o$, for $\alpha=90^o$ the form factor 
is dominated by the pair production process, except near $q^2=0$. 
At high values of the
momentum transfer the form factor is completely exhausted by the pair-term
contribution. It is worth noting that such a dominance is mainly 
due to a kinematical effect and 
appears to be fairly model independent
(see also Ref.\cite{bakker01}). 
A qualitative argument, which is applicable to the reference frame with
$q^+=\sqrt{-q^2}$ and essentially follows the
work of Sawicki\cite{sawicki}, is given in what follows. 

 Since the form factor is dimensionless, one
 can write qualitatively 
\begin{eqnarray}
F_\pi^{(I)}(q^2,\alpha)\sim \int^{P^+}_0\frac{dk^+}{P^++P^{\prime +}}
\sim \frac{P^+}{P^{\prime +}+P^+} .
\label{q1}
\end{eqnarray}
At the same time, see the discussion in Sect. II, the contribution of 
the pair production amplitude to the form factor, is roughly
\begin{eqnarray}
 F_\pi^{(II)}(q^2,\alpha)\sim \int^{P^{\prime +}}_{P^+}
\frac{dk^+}{P^++P^{\prime +}}\sim \frac{q^+}{P^{\prime +}+P^+} \ .
\label{q2}
\end{eqnarray}
The validity of such an approximation is related to the momentum cutoff  in
 the vertex
function, and therefore it is not reliable for large values of $q^+$. 
Recalling that for $\alpha=90^o$
\begin{eqnarray}
 P^+=\sqrt{m^2_\pi+ \left(\frac{q^+}{2}\right)^2}-\frac{q^+}{2} \, 
\label{q0}
\end{eqnarray}
the  estimate  of the ratio of the above contributions is given by
\begin{eqnarray}
\frac{F_\pi^{(I)}(q^2,\alpha=90^o)}{F_\pi^{(II)}(q^2,\alpha=90^o)} 
\sim \sqrt{\left(\frac{m_\pi}{q^+}\right)^2+\frac14}-\frac12 \ . 
\label{q3}
\end{eqnarray}
At the qualitative level, we can roughly say  from Eq. (\ref{q3}) that  the 
two contributions to the form factor 
are expected to have about the same magnitude when $-q^2=-q^2_{(I/II)}= m^2_\pi/2$, 
which gives
for the pion $-q^2_{(I/II)}=$   0.01 (GeV/c)$^2$. Our model 
calculation yields 
$-q^2_{(I/II)}=$ 0.03 (GeV/c)$^2$. In 
Ref.\cite{bakker01}, for the kaon it was shown that the pair term contribution
becomes dominant for  $-q^2_{(I/II)}\approx$ 0.2 (GeV/c)$^2$, while 
our estimate (Eq. (\ref{q3})) gives $-q^2_{(I/II)}~=~m^2_K/2~=$  0.13 (GeV/c)$^2$. 
Therefore, one could argue that our crude estimate is able to give 
a momentum scale 
where 
sizeable effects  due to the pair production term  are expected for 
$\alpha=90^o$.

The contributions of the instantaneous part of the Dirac propagator to
 $F_\pi^{(I)}(q^2,\alpha)$ and $F_\pi^{(II)}(q^2,\alpha)$, called
$F_{\pi \ inst}^{(I)}(q^2,\alpha)$ and  $F_{\pi \ inst}^{(II)}(q^2,\alpha)$,
respectively, are also shown in Fig. 2 for $\alpha=90^o$. 
One can physically understand why in $F_\pi^{(II)}(q^2,\alpha)$
the instantaneous part is important and
dominates at high momentum transfers. The 
interpretation
is the following: in principle, see Fig. 1 (b), the spectator quark can be exchanged 
between  incoming and outgoing pion
at a given instant $x^+$, while the pair of quark and
antiquark has been produced by the virtual photon at an earlier stage.
As the magnitude of the momentum $q^-(=-q^+)$ increases,
 the time fluctuation for the virtual
process decreases and favours the instantaneous exchange of the spectator
quark between the initial and final pion. In fact, Fig. 2 shows the
dominance of $F_{\pi \ inst}^{(II)}(q^2,\alpha)$ in the pion form factor
as the momentum transfer increases. It is worth noting that  the value 
of $F_{\pi \ inst}^{(I)}(q^2,\alpha)$ is nonzero 
because of the specific analytic structure of the vertex.
As a matter of fact, $F_{\pi \ inst}^{(I)}(q^2,\alpha)$ is nonzero, 
because of the presence of a pole in Eq. (\ref{FI}) 
at $k^-=(k^2_\perp+m^2_R)/k^+$ in  
the vertex function $\Lambda$,  Eq. (\ref{llf}), when 
one choses to close the contour for the Cauchy integration
in the lower complex semi-plane of $k^-$.

 In Fig. 3,   the results for 
the various contributions to the pion form factor for $-q^2=1$ (GeV/c)$^2$
as a function of the angle $\alpha$ are shown. For increasing angles, the
form factor changes  smoothly from valence 
to pair-term or nonvalence dominance.

In Fig. 4, we compare the results 
for $F^{({I})}_\pi$, Eq. (\ref{FI}), and $F^{(WF)}_\pi$, Eq. (\ref{Fwf}), 
where the light-front valence component of the model, defined according to 
Eq. (\ref{wf2}) is used. The absolute normalization of the valence component
$\eta$, i.e., the probability of the $q\overline q$ Fock-state component 
in the pion,  is calculated to be $\eta=0.77$, differently from 
 the nonsymmetric regulator model  of  
Ref. \cite{pipach}, where $\eta=1$. The symmetric form of the vertex 
implies contributions from many poles 
to the form factor and the presence of many poles makes the valence component 
comparatively smaller than in the nonsymmetric case.
In a previous work on DIS, based on a wave function contribution only,
\cite{tob92}
a renormalization   around 0.5 - 0.75 was necessary to
fit the data. 
To compare  $F^{({I})}_\pi$ and $F^{(WF)}_\pi$,
 we have arbitrarily normalized $F^{(WF)}_\pi(0)$ to 1. 
As shown in Fig. 4, the momentum behaviour
of $F^{(I)}_\pi$ and $F^{(WF)}_\pi$ is almost the same, 
independently of the reference frame, for $\alpha$ between
$0^o$ and $90^o$. 
This means  that our kinematical
argument, about the suppression of $F^{(I)}_\pi(q^2,\alpha)$ with respect
to the full form factor in the $q^+=\sqrt{-q^2}$ frame, could be
extended to $F^{(WF)}_\pi(q^2,\alpha)$ as well. 

Once $F^{(WF)}_\pi(0)$ is arbitrarily normalized to 1, namely $C \rightarrow
C/\sqrt{\eta}$ in Eq. (\ref{nwf}), 
a good description of the form factor data for $\alpha=0^o$ is achieved, 
as one can deduce from
Fig. 4,  but 
$f^{(WF)}_\pi = 105$ MeV, which overestimates the experimental value, since
 $f^{(WF)}_\pi=f_\pi/\sqrt{\eta}$.
The value of $f^{(WF)}_\pi$ is similar to the  one found
for nonsymmetric models, 
once the form factor below 2 (GeV/c)$^2$ is fitted (see e. g. \cite{pipach}).

As discussed in Sect. III, $F^{(WF)}_\pi$ can be formally related to the form
factor obtained within FFHD,   where only  the valence 
contribution appears.
Since the momentum behaviour
of $F^{(I)}_\pi$ and $F^{(WF)}_\pi$ is almost the same,
  then one can argue that 
$F^{FFHD}_\pi$ should correspond to the contribution of $F^{(I)}_\pi $. 
This
means that $F^{FFHD}_\pi$ calculated in  the Drell-Yan frame could 
represent a good effective
approach for evaluating the form factor of the pion.
Let us note that in FFHD 
 the pair term
 can appear only as a contribution from     two-body currents.

\subsection{Weakly relativistic systems}

In order to investigate the sensitivity of the pair term upon the dynamical
scale of the composite system, we consider a case sharply
different from the pion one. We adjust our model to the deuteron scale, 
and use for the mass of the system the value $m_D=1.874$ GeV and 
for the mass of the constituents  $m=0.938$ GeV. 
We adopt a regulator mass of $m_R=1.1$ GeV that gives 
a mean square radius of 3.25 fm$^2$ comparable to the 
difference, $r^2_{D,exp}-r^2_{p,exp}$, between the experimental
values of the deuteron and proton mean square radii.
In Fig. 5, we show the results of the form factor calculations 
for our mock deuteron.
 According to our qualitative kinematical 
estimate 
given in the previous subsection 
(see Eq. (\ref{q3}),   with $m_D$ replacing $m_\pi$), at $-q^2 \sim $ 2 (GeV/c)$^2$
the pair term in the Breit frame with $q^+=\sqrt{-q^2}$
 is expected to be as large as 
$F^{(I)}$.
As a matter of fact, at this momentum transfer, the pair term is only 
about 15\% of the form factor,
as shown in Fig. 5. Indeed, the kinematical estimate does not work for a system
with a rapidly decreasing momentum distribution. Moreover, we observe that
the instantaneous term of the Dirac propagator gives a small
contribution to the pair term, as well as to $F^{(I)}(q^2,\alpha=90^o)$. 
This fact is
related to the basically nonrelativistic nature of the costituents in our
mock deuteron. As illustrated in Fig. 6, the dimensionless
product of $k^2_\perp$ times the transverse 
probability density, Eq. (\ref{prob1}), is peaked at $k_\perp=0.06~m$
for our model. For the sake of comparison, in Fig. 6 it is also shown the
  transverse 
probability density for an actual deuteron; in particular the density has been
calculated from the deuteron wave function corresponding to  the Reid soft-core model of the 
deuteron\cite{reid}. Remarkably, the same overall behaviour  has been obtained
in both cases.
On the contrary the pion
is a strongly relativistic system. For our pion model $k^2_\perp f(k_\perp)$
is peaked at $k_\perp \simeq m$, while for the FFHD model of 
Refs.\cite{salme,card} the peak occurs around $k_\perp=2.5m$. 
The FFHD model of Refs.\cite{salme,card}
 is based on a mass operator with the
effective interaction of Ref.\cite{gi}, which includes one-gluon-exchange
and linear confinement terms. The pion model from this mass operator, in the
$q^+=0$ frame, has a charge radius of 0.46 fm for pointlike constituent quarks. 
This value largely explains why
the position of the peak appears at an higher value of the
transverse momentum with respect to our model, which has a radius
of 0.74 fm. 

In summary, for a weakly relativistic system with spin equal to 0, 
the valence contribution to
the form factor
yields a good approximation of the whole form factor up to $-q^2=2~(GeV/c)^2$ in
any reference frame. Therefore, one can argue that  the evaluation 
of $F^{FFHD}$ can be considered a reliable approximation to the full form 
factor in this case.

\subsection{Pair-term contribution and FFHD estimate}

In order to study  in more detail the pair-term contribution, we show in Fig. 7
the ratio between 
$F^{(II)}(q^2,\alpha=90^o)$ 
and the full form factor.  We  
compare the results of our covariant model 
\begin{eqnarray}
R^{(COV)}=\frac{ F^{(II)}(q^2,\alpha=90^o)}{F(q^2)}
\label{ratio}
\end{eqnarray}
with the kinematical estimate, Eqs. (\ref{q1}) and  
(\ref{q2}), given by
\begin{eqnarray}
 R^{(KIN)}=\frac{q^+}{q^++P^+} \ .
\label{ratiok}
\end{eqnarray}
This kinematical estimate of the pair term works fairly well for the pion, while it 
overestimates  the contribution of $F^{(II)} $ for the
weakly relativistic system, due to the rapid fall-off of the momentum
distribution which makes  Eq. (\ref{q2}) unreliable.  In Fig. 7, we also 
report the following ratio 
\begin{eqnarray}
R^{(FFHD)}=
\frac{F_\pi^{(DY)}(q^2) -F_\pi^{(LPS)}(q^2) }{F_\pi^{(DY)}(q^2)} \ .
\label{rffhd}
\end{eqnarray}
where $F_\pi^{(DY)}(q^2)$ and 
$F_\pi^{(LPS)}(q^2)$ are the pion form factor evaluated 
in the Drell-Yan frame and in the $q^+=\sqrt{-q^2}$ frame,
respectively, within the FFHD model of Ref.\cite{card}, using pointlike
constituent quarks and the one-body current. 
Such a ratio could yield insight on the
relevance in FFHD of two-body current contributions  corresponding to the pair term. 
Indeed,  $F_\pi^{(DY)}(q^2)$ represents an estimate
of $F_\pi^{(I)}(q^2,\alpha=0^o)=F_\pi(q^2)$, while $F_\pi^{(LPS)}(q^2)$ should describe 
$F_\pi^{(I)}(q^2,\alpha=90^o)$ (cf. comments to Fig. 4) and therefore one can argue
that $R^{(FFHD)} \sim R^{(COV)}\sim  R^{(KIN)}$. 
As a matter of fact, the ratio $R^{(FFHD)}$  qualitatively agrees with 
the results of our covariant model 
and with the model-independent kinematical estimate (see Fig. 7). Thus, the
suppression of $F_\pi^{FFHD}$ in the  $q^+ =\sqrt{-q^2}$ frame
with respect to the Drell-Yan frame   can be
explained in terms of a missing contribution in the current, 
that we could identify with two-body current contributions
related to the  pair term. 

For the sake of completeness, $R^{(COV)}$ and   $R^{(KIN)}$  are also shown in 
Fig. 7 for a weakly
relativistic sytem. Since 
$R^{(KIN)} >> R^{(COV)}$,  one can rely only on model calculations
for inferring the dominance of the valence contribution 
with respect to  the nonvalence, over a 
wide range of momentum transfer.
 We have also evaluated $R^{(FFHD)}$ for  our mock deuteron and we have 
found $R^{(FFHD)} \sim R^{(COV)}$. This result confirms the small effect of the
pair term for a spin-0 weakly relativistic system.

\section{Conclusion}

Within a covariant model with a  vertex function which is symmetric
in the momentum space
of the constituents, we have calculated the electromagnetic form factor
for a two-fermion, pion-like system.
Such a model has allowed us to perform a detailed analysis of the pair-term 
contribution
both changing the orientation of the Breit frame, where the form factor is 
calculated,
and the dynamical scale of the system. 

We have used a pseudoscalar coupling for the pion-$q\bar{q}$ vertex 
and
a current  for the point-like constituents in impulse approximation. 
It is worth noting 
that in our approach the current of the whole system is conserved. 
We have adopted for the first time a nonconstant, symmetric  ansatz for 
the Bethe-Salpeter amplitude, and
 we have obtained 
the valence component
of the pion state by projecting out  the Bethe-Salpeter amplitude
on the $x^+=0$ hypersurface. The symmetric vertex function implies a valence
wave function with the same property. In this way we have overcomed 
some previous conceptual and phenomenological difficulties related to the  
use of a nonsymmetric regulator for describing the pion. As a matter   
of fact, 
the form factor and the weak decay constant cannot be 
simultaneously reproduced  with a nonsymmetric 
vertex function (see, e. g., \cite{pipach}). 

In our covariant model for the pion, the two free parameters,
 the constituent quark mass $(m)$
and the regulator mass $(m_R)$, have been fixed as follows:
i) the constituent quark mass was
chosen as $m=0.220$ GeV from the meson spectroscopy \cite{gi},
ii) the regulator mass $m_{R}=0.6$ GeV was found by fitting the
experimental weak decay constant, $f^{exp}_\pi=92.4$ MeV.
As a consequence, the form factor obtained in our covariant model  
nicely agrees with the new data for the pion form 
factor \cite{tj}. It should be pointed out  that, if we wish to 
reproduce simultaneously $f_\pi$ and the form factor data with our 
approach, the constituent quark mass should be chosen 
 in the range between 0.2 and 0.3 GeV.
Another interesting outcome of our symmetric model is that the probability of the
pion valence component, $\eta$, is less than 1  (about $75\%$), at variance with 
previous covariant calculations \cite{pipach,bakker01} where  $\eta~=~1$.
The em form factor of the pion-like system has been decomposed in two
contributions, with the second one given by the pair-diagram term.
The separation of the covariant result in such contributions 
is unique and does not depend on any particular dynamical model used to
generate our choice of BS vertex, as long as the four-dimensional
impulse approximation is adopted to calculate the em current (see Eq. (\ref{jmu})
).

Following Ref. \cite{bakker01}, but with a nonconstant, symmetric vertex function,
we have investigated the frame dependence of the pair-term contribution
to the form factor of pion-like systems. Our conclusions strengthen and
generalize the conclusions of Ref. \cite{bakker01}. In particular, after 
introducing 
a transverse momentum distribution to distinguish between strongly and
weakly relativistic systems, we have
also analyzed  the form factor for the second case.

In general the  magnitude of the pair-term contribution in the electromagnetic
current depends on the frame, because a dynamical transformation change
the relative weight of   valence and nonvalence  contributions in the Fock state
and then in the
form factor, while the sum, that is an invariant  quantity, does not.
For strongly relativistic  systems, as the pion, the effect is dramatic: in  
the Breit frame where $q^+=\sqrt{-q^2}$ 
the form factor is largely dominated by the pair diagram for $-q^2 \ne 0$,
while adjusting the model
to the deuteron scale (weakly relativistic system), 
the pair term becomes negligible at low momentum transfer and only
contributes by  30\% at $-q^2=4~(GeV/c)^2$. 

The rapid fall-off of the valence component contribution to the 
form factor of the pion  
for purely longitudinal momentum transfers, $ q^+=\sqrt{-q^2}$,
is a  feature   which can be understood also by
a general, crude kinematical estimate. As matter of fact, the ratio between
the pair contribution and the full form factor is roughly given by
(see also, e. g., \cite{sawicki})
$$R^{KIN}=\frac{q^+}{\sqrt{m^2_\pi+(\frac{q^+}{2})^2}+\frac{q^+}{2}} \ .$$
From such an estimate one can have some insight on  the role played by 
the mass of the composite system, in determining
the onset of the importance of the pair diagram in the 
form factor. 

Finally, we have compared the results of our covariant model
for the pion form factor
to the ones developed within the Front Form Hamiltonian Dynamics,
based on one-body current operators. Such a comparison  has shown that some
general features are shared. In particular,  
the    fall-off of the valence
contribution   in the  $q^+=\sqrt{-q^2}$ frame faster than the one in the
$q^+=0$ frame 
can be recovered in the FFHD approach. 
It should be pointed out that the FFHD calculations 
do not include the contribution of the pair term to the form factor,
which should appear as a two-body current contribution.

In the case of the deuteron scale (weakly relativistic system), 
the relevance of the pair-term contribution is shifted towards higher
values of the momentum transfer (2-4 (GeV/c)$^2$). Therefore, 
these encouraging results of our exploratory analysis at the deuteron scale 
urge  studies 
which take into account the vector nature
of the deuteron.

The investigation of the pair-term contribution carried out
within the covariant model of the present work  could
give a first suggestion for an explicit form of the 
two-body current contribution to be used
within phenomenological FFHD approaches.
A detailed analysis of the light-front pair term in a given
field theoretical model can be done using the quasi-potential
approach of Refs. \cite{sales1,sales2}. 
Within this approach, the light-front 
pair diagram for
nonvanishing $q^+$  will come from the contribution of 
a two-body current, which can be formally derived in a consistent way. 
Numerical investigations of a dynamical model, featuring the 
mentioned properties and suitable for applications  to hadrons,     
will be presented elsewhere.

\section*{Acknowledgments}
We gratefully acknowledge Felix Lev for interesting discussions about
the symmetry properties of the vertex function.
This work was partially supported by
the Brazilian agencies CNPq and 
FAPESP and by Ministero della Ricerca Scientifica e Tecnologica. 
T.F. acknowledges the hospitality of the
Dipartimento di Fisica, Universit\`a di Roma "Tor Vergata" and 
of Istituto
Nazionale di Fisica Nucleare, Sezione Tor Vergata and Sezione Roma I.

\newpage

\appendix
\section{Analytic Integration on $k^-$ in $F^{(I)}_{\pi}$}

In this Appendix, we show in detail how to perform  the contour 
integration on $k^-$ 
in Eq. (\ref{FI}) for $F^{(I)}_{\pi}(q^2,\alpha)$. This quantity can be  rewritten as
follows
\begin{eqnarray}
F_\pi^{(I)}(q^2,\alpha)= F_\pi^{(I)a}+F_\pi^{(I)b}+F_\pi^{(I)c}+F_\pi^{(I)d}
\ .
\label{a1}
\end{eqnarray}

The first term in Eq. (\ref{a1}), $ F_\pi^{(I)a}$, is 
\begin{eqnarray}
F_\pi^{(I)a}&=&- \frac{\imath}{2\pi} N
 \int \frac{d^{2} k_{\perp} d k^{+} d k^-\theta (k^+)\theta(P^+-k^+)}
{(k^+)^3(P^+-k^+) (P^{^{\prime}+}-k^+)} Tr[{\cal O}^+(k^-)]
\nonumber \\ &\times& 
\frac{1}
{ (k^- - k^-_{on}+\imath \epsilon)
(P^- - k^- -(P-k)^-_{on}+ \imath\epsilon)}
\nonumber \\ &\times& 
\frac{1}
{(P^{\prime -} - k^- -(P^\prime-k)^-_{on}+\imath \epsilon)
(k^- - k^-_R+\imath \epsilon)^2} \ ,
\label{a2}
\end{eqnarray} 
where we explicitly wrote the dependence upon $k^-$ 
in $Tr[{\cal O}^+(k^-)]$ and 
\begin{eqnarray}
k^-_R=\frac{k_{\perp}^{2}+m^2_R}{k^+} \ ,
\label{a3}
\end{eqnarray} 
The on-energy-shell values of the individual momenta 
$k^-_{on}$, $(P-k)^-_{on}$ and $(P^\prime-k)^-_{on}$ are given by 
Eq. (\ref{onek}).
The normalization factor is
\begin{eqnarray}
N=\frac{m^2 C^2}{(P^{+}+P^{\prime +})f^2_\pi} 
\frac{N_c}{(2\pi)^3} 
\label{a4}
\end{eqnarray} 

For the sake of algebraic simplicity the contour of integration 
in Eq. (\ref{a2}) can be closed in the upper complex semi-plane of $k^-$, where
only the poles $k^-_{(1)}$ and $k^-_{(3)}$, given by Eqs. (\ref{k1})
and (\ref{k3}), respectively, are present. By evaluating  
the residues of  the integrand in Eq. (\ref{a2}), one has
\begin{eqnarray}
F_\pi^{(I)a}&=&-N
 \int \frac{d^{2} k_{\perp} d k^{+} \theta (k^+)\theta(P^+-k^+)}
{(k^+)^3(P^+-k^+) (P^{^{\prime}+}-k^+)}
\frac{1}{(P^{\prime -} - P^{-} +(P-k)^-_{on} -(P^\prime-k)^-_{on})}
\nonumber \\ &\times& 
\left[ 
\frac{Tr[{\cal O}^+(P^{-} -(P-k)^-_{on})]}
{(P^{-} -(P-k)^-_{on}- k^-_{on})
(P^{-} -(P-k)^-_{on} - k^-_R)^2} \right.
\nonumber \\ &-& 
\left.
\frac{Tr[{\cal O}^+(P^{\prime-} -(P^\prime-k)^-_{on})]}
{(P^{\prime -} -(P^\prime-k)^-_{on}- k^-_{on})
(P^{\prime -} -(P^\prime-k)^-_{on} - k^-_R)^2} \right]
\ .
\label{a5}
\end{eqnarray} 

The second term in Eq. (\ref{a1}), $ F_\pi^{(I)b}$, is given by
\begin{eqnarray}
F_\pi^{(I)b}&=&- \frac{\imath}{2\pi} N
 \int \frac{d^{2} k_{\perp} d k^{+} d k^-\theta (k^+)\theta(P^+-k^+)}
{(k^+)^2(P^+-k^+) (P^{^{\prime}+}-k^+)^2} Tr[{\cal O}^+(k^-)]
\nonumber \\ &\times& 
\frac{1}
{ (k^- - k^-_{on}+\imath \epsilon)
(P^- - k^- -(P-k)^-_{on}+ \imath\epsilon)
(P^{\prime -} - k^- -(P^\prime-k)^-_{on}+\imath \epsilon)}
\nonumber \\ &\times& 
\frac{1}
{(k^- - k^-_R+\imath \epsilon)
(P^{\prime -} - k^- -(P^\prime-k)^-_R+\imath \epsilon)} \ ,
\label{a6}
\end{eqnarray} 
where 
\begin{eqnarray}
(P^\prime-k)^-_R=\frac{(P^\prime-k)_{\perp}^{2}+m^2_R}{P^{\prime +}-k^+} \ .
\label{a7}
\end{eqnarray} 
In Eq. (\ref{a6}), the integration contour  is closed
in the lower complex semi-plane of $k^-$, where
the poles $k^-_{(5)}=k^-_{on}-\imath \epsilon$ and 
$k^-_{(6)}=k^-_R-\imath \epsilon$ are present. By calculating
the residues of the integrand in Eq. (\ref{a6}), one gets
\begin{eqnarray}
F_\pi^{(I)b}&=&-N
 \int \frac{d^{2} k_{\perp} d k^{+} \theta (k^+)\theta(P^+-k^+)}
{(k^+)^2(P^+-k^+) (P^{^{\prime}+}-k^+)^2}
Tr[{\cal O}^+(k^-_{on})]
\nonumber \\ &\times& 
 \frac{1}{(P^{-} -(P-k)^-_{on}- k^-_{on})
(P^{\prime -} -(P^\prime - k)^-_{on} - k^-_{on})} 
\nonumber \\ &\times& 
\frac{1}
{(k^-_{on}-k^-_R)
(P^{\prime -}  - k^-_{on}-(P^\prime-k)^-_R)} \ \
+\left[ k^-_{on}\leftrightarrow k^-_R\right] \ .
\label{a8}
\end{eqnarray} 

The third term in Eq. (\ref{a1}), $ F_\pi^{(I)c}$, is given by
\begin{eqnarray}
F_\pi^{(I)c}&=&- \frac{\imath}{2\pi} N
 \int \frac{d^{2} k_{\perp} d k^{+} d k^-\theta (k^+)\theta(P^+-k^+)}
{(k^+)^2(P^+-k^+) (P^{^{\prime}+}-k^+)^2} Tr[{\cal O}^+(k^-)]
\nonumber \\ &\times& 
\frac{1}
{ (k^- - k^-_{on}+\imath \epsilon)
(P^- - k^- -(P-k)^-_{on}+ \imath\epsilon)
(P^{\prime -} - k^- -(P^\prime-k)^-_{on}+\imath \epsilon)}
\nonumber \\ &\times& 
\frac{1}
{(k^- - k^-_R+\imath \epsilon)
(P^{-} - k^- -(P-k)^-_R+\imath \epsilon)} \ .
\label{a9}
\end{eqnarray} 
Eq. (\ref{a9}) is identical to Eq. (\ref{a6}) with
 $P^\prime\leftrightarrow P$. Consequently we can write 
\begin{eqnarray}
F_\pi^{(I)c}=F_\pi^{(I)b}\left[P^\prime\leftrightarrow P\right] \ .
\label{a10}
\end{eqnarray} 

The last term in Eq. (\ref{a1}), $ F_\pi^{(I)d}$, that represents the full
  $ F_\pi^{(I)}$ for the nonsymmetric model of Ref. \cite{pipach}, is given by
\begin{eqnarray}
F_\pi^{(I)d}&=&- \frac{\imath}{2\pi} N
 \int \frac{d^{2} k_{\perp} d k^{+} d k^-\theta (k^+)\theta(P^+-k^+)}
{k^+(P^+-k^+)^2 (P^{^{\prime}+}-k^+)^2} Tr[{\cal O}^+(k^-)]
\nonumber \\ &\times& 
\frac{1}
{ (k^- - k^-_{on}+\imath \epsilon)
(P^- - k^- -(P-k)^-_{on}+ \imath\epsilon)
(P^{\prime -} - k^- -(P^\prime-k)^-_{on}+\imath \epsilon)}
\nonumber \\ &\times& 
\frac{1}
{(P^- - k^- -(P-k)^-_R+ \imath\epsilon)
(P^{\prime -} - k^- -(P^\prime-k)^-_R+\imath \epsilon)} \ ,
\label{a11}
\end{eqnarray} 
where 
\begin{eqnarray}
(P-k)^-_R=\frac{(P-k)_{\perp}^{2}+m^2_R}
{P^{ +}-k^+} \ .
\label{a12}
\end{eqnarray} 
In  Eq. (\ref{a11}), the integration contour can be closed
 in the lower complex semi-plane of $k^-$, where only
the pole $k^-_{(5)}=k^-_{on}-\imath \epsilon$ is present. By evaluating 
the residue of the integrand in Eq. (\ref{a11}), one has
\begin{eqnarray}
F_\pi^{(I)d}&=&- N
 \int \frac{d^{2} k_{\perp} d k^{+} \theta (k^+)\theta(P^+-k^+)}
{k^+(P^+-k^+)^2 (P^{^{\prime}+}-k^+)^2} Tr[{\cal O}^+(k^-_{on})]
\nonumber \\ &\times& 
\frac{1}
{ (P^- - k^-_{on} -(P-k)^-_{on})
(P^{\prime -} - k^-_{on} -(P^\prime-k)^-_{on})}
\nonumber \\ &\times& 
\frac{1}
{(P^- - k^-_{on} -(P-k)^-_R)
(P^{\prime -} - k^-_{on} -(P^\prime-k)^-_R)} \ .
\label{a13}
\end{eqnarray} 
\newpage

\section{Analytic Integration on $k^-$ in $F^{(II)}_\pi$}

To perform  the contour integration    on $k^-$ 
in Eq. (\ref{FII}) for the pair term, $F^{(II)}_\pi(q^2,\alpha)$, we rewrite such a
contribution as follows
\begin{eqnarray}
F_\pi^{(II)}(q^2,\alpha)= F_\pi^{(II)a}+F_\pi^{(II)b}+F_\pi^{(II)c}+F_\pi^{(II)d}
\ .
\label{b1}
\end{eqnarray}
 All the terms in Eq. (\ref{b1}) can be evaluated by closing 
 the integration contour  
  in the upper complex semi-plane of $k^-$.
The first term in Eq. (\ref{b1}), $ F_\pi^{(II)a}$ is
\begin{eqnarray}
F_\pi^{(II)a}&=&- \frac{\imath}{2\pi} N
 \int \frac{d^{2} k_{\perp} d k^{+} d k^-
\theta (k^+-P^+)\theta(P^{\prime +}-k^+)}
{(k^+)^3(P^+-k^+) (P^{^{\prime}+}-k^+)} Tr[{\cal O}^+(k^-)]
\nonumber \\ &\times& 
\frac{1}
{ (k^- - k^-_{on}+\imath \epsilon)
(P^- - k^- -(P-k)^-_{on}- \imath\epsilon)}
\nonumber \\ &\times& 
\frac{1}
{(P^{\prime -} - k^- -(P^\prime-k)^-_{on}+\imath \epsilon)
(k^- - k^-_R+\imath \epsilon)^2} \ ,
\label{b2}
\end{eqnarray} 
where the on-energy-shell values of the individual momenta
$k^-_{on}$, $(P-k)^-_{on}$ and $(P^\prime-k)^-_{on}$ are again given by 
Eq. (\ref{onek}) and $k^-_R$ is defined by Eq. (\ref{a3}) .
The normalization factor is the same as in Eq. (\ref{a4}).

In Eq. (\ref{b2}), 
only the pole  $k^-_{(3)}$ given by Eq. (\ref{k3}) is present
in the upper complex semi-plane of $k^-$. By evaluating
the residue of the integrand in Eq. (\ref{b1}), one gets
\begin{eqnarray}
F_\pi^{(II)a}&=&-N
 \int \frac{d^{2} k_{\perp} d k^{+} 
\theta (k^+-P^+)\theta(P^{\prime +}-k^+)}
{(k^+)^3(P^+-k^+) (P^{^{\prime}+}-k^+)}
\frac{Tr[{\cal O}^+(P^{\prime-} -(P^\prime-k)^-_{on})]}
{(P^{\prime-} -(P^\prime-k)^-_{on}-k^-_{on})}
\nonumber \\ &\times& 
\frac{1}{(P^{-} - P^{\prime -} -(P-k)^-_{on} +(P^\prime-k)^-_{on})
(P^{\prime -} -(P^\prime-k)^-_{on}- k^-_{R})^2}
\ .
\label{b3}
\end{eqnarray} 

The second term in  Eq. (\ref{b1}), $ F_\pi^{(II)b}$, is given by
\begin{eqnarray}
F_\pi^{(II)b}&=&- \frac{\imath}{2\pi} N
 \int \frac{d^{2} k_{\perp} d k^{+} d k^-
\theta (k^+-P^+)\theta(P^{\prime +}-k^+)}
{(k^+)^2(P^+-k^+) (P^{^{\prime}+}-k^+)^2} Tr[{\cal O}^+(k^-)]
\nonumber \\ &\times& 
\frac{1}
{ (k^- - k^-_{on}+\imath \epsilon)
(P^- - k^- -(P-k)^-_{on}- \imath\epsilon)
(P^{\prime -} - k^- -(P^\prime-k)^-_{on}+\imath \epsilon)}
\nonumber \\ &\times& 
\frac{1}
{(k^- - k^-_R+\imath \epsilon)
(P^{\prime -} - k^- -(P^\prime-k)^-_R+\imath \epsilon)} \ .
\label{b4}
\end{eqnarray} 
In Eq. (\ref{b4}), 
the poles $k^-_{(3)}$, Eq. (\ref{k3}), and 
$k^-_{(4)}$, Eq. (\ref{k4}), contribute. By calculating
the residues of the integrand in Eq. (\ref{b4}), one has
\begin{eqnarray}
F_\pi^{(II)b}&=&-N
 \int \frac{d^{2} k_{\perp} d k^{+} 
\theta (k^+-P^+)\theta(P^{\prime +}-k^+)}
{(k^+)^2(P^+-k^+) (P^{^{\prime}+}-k^+)^2}
Tr[{\cal O}^+(P^{\prime -} -(P^\prime-k)^-_{on})]
\nonumber \\ &\times& 
 \frac{1}
{((P^\prime-k)^-_{on}-(P^\prime-k)^-_{R})
(P^{\prime -} -(P^\prime-k)^-_{on}- k^-_{on})}
\nonumber \\ &\times& 
\frac{1}{(P^{\prime -} -(P^\prime-k)^-_{on}- k^-_{R})
(P^--P^{\prime -} +(P^\prime - k)^-_{on} -(P - k)^-_{on} )}
\nonumber \\ &+& 
\left[ (P^\prime-k)^-_{on}\leftrightarrow(P^\prime-k)^-_{R}\right]
 \ .
\label{b5}
\end{eqnarray} 

The third term in Eq. (\ref{b1}), $ F_\pi^{(II)c}$, is 
\begin{eqnarray}
F_\pi^{(II)c}&=&- \frac{\imath}{2\pi} N
 \int \frac{d^{2} k_{\perp} d k^{+} d k^-
\theta (k^+-P^+)\theta(P^{\prime +}-k^+)}
{(k^+)^2(P^+-k^+) (P^{^{\prime}+}-k^+)^2} Tr[{\cal O}^+(k^-)]
\nonumber \\ &\times& 
\frac{1}
{ (k^- - k^-_{on}+\imath \epsilon)
(P^- - k^- -(P-k)^-_{on}- \imath\epsilon)
(P^{\prime -} - k^- -(P^\prime-k)^-_{on}+\imath \epsilon)}
\nonumber \\ &\times& 
\frac{1}
{(k^- - k^-_R+\imath \epsilon)
(P^{-} - k^- -(P-k)^-_R-\imath \epsilon)} \ .
\label{b6}
\end{eqnarray} 
In Eq. (\ref{b6})  the  only pole
of the integrand in the upper complex semi-plane of $k^-$ is $k^-_{(3)}$,
given by Eq. (\ref{k3}). By computing the residue, the
result is
\begin{eqnarray}
F_\pi^{(II)c}&=&- N
 \int \frac{d^{2} k_{\perp} d k^{+} 
\theta (k^+-P^+)\theta(P^{\prime +}-k^+)}
{(k^+)^2(P^+-k^+) (P^{^{\prime}+}-k^+)^2} 
Tr[{\cal O}^+(P^{\prime -}-(P^\prime-k)^-_{on})]
\nonumber \\ &\times& 
\frac{1}
{ (P^{\prime -}-(P^\prime-k)^-_{on} - k^-_{on})
(P^- - P^{\prime -}+(P^\prime-k)^-_{on} -(P-k)^-_{on})
}
\nonumber \\ &\times& 
\frac{1}
{(P^{\prime -}-(P^\prime-k)^-_{on} - k^-_R)
(P^{-} - P^{\prime -}+(P^\prime-k)^-_{on} -(P-k)^-_R)} \ .
\label{b7}
\end{eqnarray} 

The last term is Eq. (\ref{b1}), $ F_\pi^{(II)d}$, is given by
\begin{eqnarray}
F_\pi^{(II)d}&=&- \frac{\imath}{2\pi} N
 \int \frac{d^{2} k_{\perp} d k^{+} d k^-
\theta (k^+-P^+)\theta(P^{\prime +}-k^+)}
{k^+(P^+-k^+)^2 (P^{^{\prime}+}-k^+)^2} Tr[{\cal O}^+(k^-)]
\nonumber \\ &\times& 
\frac{1}
{ (k^- - k^-_{on}+\imath \epsilon)
(P^- - k^- -(P-k)^-_{on}- \imath\epsilon)
(P^{\prime -} - k^- -(P^\prime-k)^-_{on}+\imath \epsilon)}
\nonumber \\ &\times& 
\frac{1}
{(P^- - k^- -(P-k)^-_R- \imath\epsilon)
~(P^{\prime -} - k^- -(P^\prime-k)^-_R+\imath \epsilon)} \ .
\label{b8}
\end{eqnarray} 
In Eq. (\ref{b8})  
the poles $k^-_{(3)}$, Eq. (\ref{k3}), and $k^-_{(4)}$,
Eq. (\ref{k4}), contribute. By calculating
the residues of the integrand of Eq. (\ref{b8}), we obtain
\begin{eqnarray}
F_\pi^{(II)d}&=&-  N
 \int \frac{d^{2} k_{\perp} d k^{+} 
\theta (k^+-P^+)\theta(P^{\prime +}-k^+)}
{k^+(P^+-k^+)^2 (P^{^{\prime}+}-k^+)^2} 
Tr[{\cal O}^+(P^{\prime -}-(P^\prime-k)^-_{on})]
\nonumber \\ &\times& 
\frac{1}
{ (P^{\prime -}-(P^\prime-k)^-_{on} - k^-_{on})
~(P^- - P^{\prime -}+(P^\prime-k)^-_{on} -(P-k)^-_{on})
}
\nonumber \\ &\times& 
\frac{1}
{(P^- - P^{\prime -}+(P^\prime-k)^-_{on} -(P-k)^-_R)
~((P^\prime-k)^-_{on} -(P^\prime-k)^-_R)}
\nonumber \\ &+& 
\left[(P^\prime-k)^-_{on} \leftrightarrow (P^\prime-k)^-_{R}\right]
 \ .
\label{b9}
\end{eqnarray}


\newpage
\centerline{FIGURE CAPTIONS}
\medskip 

{\bf Fig. 1}.  Light-front time-ordered diagrams for the current: (a) $F^{(I)}_\pi$ 
(Eq.
(\ref{FI}))
 and (b) $F^{(II)}_\pi$ (Eq.
(\ref{FII})).
\medskip 

{\bf Fig. 2}.  
The pion form factor vs $ - q^2$. The contributions to $F_\pi(q^2)$, 
evaluated in the Breit frame
with $q^+=\sqrt{-q^2}$, i.e. $\alpha=90^o$ (Eq. (\ref{alpha})),  are also shown. 
Solid line: full result; dashed line: $F_\pi^{(I)}(q^2,\alpha)$ (full result 
without
the pair term, Eq. (\ref{FI})); dotted line:
$F_\pi^{(II)}(q^2,\alpha)$ (pair term, Eq. (\ref{FII})); long-dashed line: 
$F_{\pi \ inst}^{(I)}(q^2,\alpha)$ (
instantaneous-term contribution  to $F_\pi^{(I)}(q^2,\alpha)$); short-dashed line:  
$F_{\pi \ inst}^{(II)}(q^2,\alpha)$ (
instantaneous-term contribution to the pair term).
Experimental data: Ref. \cite{tj} (full squares),
 Ref. \cite{cea} (full triangles), Ref. \cite{corn1} (empty squares),
Ref. \cite{corn2} (empty circles) and Ref.  \cite{bebek} (full circles). 
\medskip 

{\bf Fig. 3}. 
Contributions to the  pion form factor vs $\alpha$ (see Eq. (\ref{alpha})) for
$-q^2 =$ 1 (GeV/c)$^2$.
Solid line: full result; dashed line: $F_\pi^{(I)}(q^2,\alpha)$ (full result 
without
the pair term, Eq. (\ref{FI})); dotted line:
$F_\pi^{(II)}(q^2,\alpha)$ (pair term, Eq. (\ref{FII})); long-dashed line: 
$F_{\pi \ inst}^{(I)}(q^2,\alpha)$ (
instantaneous-term contribution  to $F_\pi^{(I)}(q^2,\alpha)$); short-dashed line:  
$F_{\pi \ inst}^{(II)}(q^2,\alpha)$ (instantaneous-term contribution to the pair term).
\medskip 

{\bf Fig. 4}. 
Comparison between the pion form factor calculations 
for $F_\pi^{(I)}$ (Eq. (\ref{FI})) and for $F_\pi^{(WF)}$ (Eq. (\ref{Fwf})),
at $\alpha=0^o, \ 45^o$ and 90$^o$. Solid line: $F_\pi^{(I)}$; dotted line:
$F_\pi^{(WF)}$, normalized to 1 at $q^2=0$.
\medskip 

{\bf Fig. 5}.
 Form factor for a
weakly relativistic spin-0 system  vs
$ - q^2$. The contributions to the form factor, 
evaluated in the Breit frame
with $q^+=\sqrt{-q^2}$, i.e. $\alpha=90^o$ (Eq. (\ref{alpha})),  are also shown. 
Solid line: full result; dashed line: $F_\pi^{(I)}(q^2,\alpha)$ (full result 
without
the pair term, Eq. (\ref{FI})); dotted line:
$F_\pi^{(II)}(q^2,\alpha)$ (pair term, Eq. (\ref{FII})); long-dashed line: 
$|F_{\pi \ inst}^{(I)}(q^2,\alpha)|$ (
instantaneous-term contribution  to $F_\pi^{(I)}(q^2,\alpha)$); short-dashed line:  
$F_{\pi \ inst}^{(II)}(q^2,\alpha)$ (instantaneous-term contribution to the pair term).
\medskip 

{\bf Fig. 6}.
Dimensionless product of the transverse momentum probability 
density (Eq. (\ref{prob1})) times $k^2_\perp$ vs the ratio $k_\perp/m$. Results 
for the present  pion model (solid curve with dots),
the pion model of Ref.[15] (dashed curve), a weakly relativistic system 
(solid curve) and the deuteron
from the Reid soft-core potential \cite{reid} (short-dashed curve). 

\medskip 

{\bf Fig. 7}.
Ratio between the pair term, calculated for $\alpha=90^o$,  and the full form factor vs $-q^2$.
Upper curves correspond to the pion. Solid line: covariant pion model, $R^{(COV)}$
 (Eq. (\ref{ratio}); dashed line
 with dots: FFHD ratio, $R^{(FFHD)}$
(Eq. (\ref{rffhd})); short-dashed line:
  kinematical estimate,$R^{(KIN)}$ (Eq. (\ref{ratiok})). 
Lower curves correspond to a weakly relativistic system at the deuteron scale. Solid
line: covariant result; dashed line: kinematical estimate.

\newpage

\begin{figure}

\begin{center}
\vskip 2cm
\begin{picture}(330,130)(0,0)
\Line(0,52)(60,52)
\Line(0,48)(60,48)
\Vertex(60,50){3.0}
\ArrowLine(130,100)(60,50)
\ArrowLine(60,50)(200,50)
\Vertex(200,50){3.0}
\ArrowLine(200,50)(130,100)
\Line(200,48)(260,48)
\Line(200,52)(260,52)
\Photon(130,130)(130,100){3}{6.5}
\Vertex(130,100){3.0}
\put(80,70){\makebox(0,0)[br]{$k-P$}}
\put(20,30){\makebox(0,0)[br]{$P$}}
\put(130,30){\makebox(0,0)[br]{$k$}}
\put(210,70){\makebox(0,0)[br]{$k-P^\prime$}}
\put(250,30){\makebox(0,0)[br]{$P^\prime$}}
\DashLine(100,120)(100,30){5}
\DashLine(160,120)(160,30){5}
\put(130,10){\makebox(0,0)[br]{$(a)$}}
\end{picture}
\vskip 2cm
\begin{picture}(330,130)(0,0)
\Line(0,52)(60,52)
\Line(0,48)(60,48)
\Vertex(60,50){3.0}
\ArrowLine(0,100)(60,50)
\ArrowLine(60,50)(200,50)
\Vertex(200,50){3.0}
\ArrowLine(200,50)(0,100)
\Line(200,48)(260,48)
\Line(200,52)(260,52)
\Photon(-50,120)(0,100){3}{8.5}
\Vertex(0,100){3.0}
\put(20,70){\makebox(0,0)[br]{$k-P$}}
\put(20,30){\makebox(0,0)[br]{$P$}}
\put(130,30){\makebox(0,0)[br]{$k$}}
\put(120,80){\makebox(0,0)[br]{$k-P^\prime$}}
\put(250,30){\makebox(0,0)[br]{$P^\prime$}}
\DashLine(35,120)(35,30){5}
\DashLine(160,120)(160,30){5}
\put(130,10){\makebox(0,0)[br]{$(b)$}}
\end{picture}
\end{center}
\end{figure}
\vspace{2cm}

{\bf  Fig. 1 - J. P. B. C. de Melo, T. Frederico, E. Pace and 
G.Salm\`e}

\newpage 

\begin{figure}
\psfig{bbllx=20mm,bblly=30mm,bburx=100mm,bbury=220mm,file=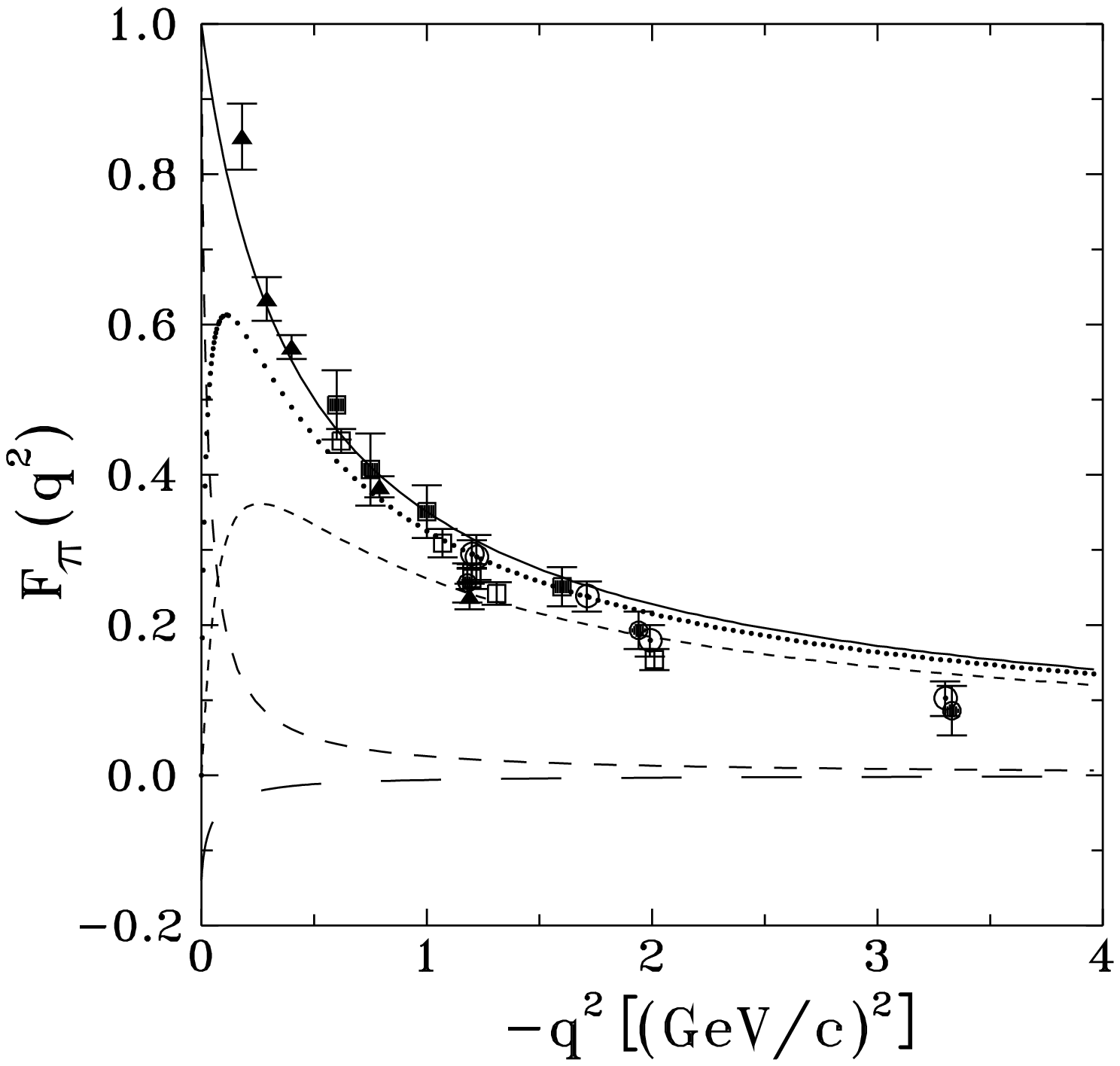}
\end{figure}
\vspace{2cm}

{\bf Fig. 2 - J. P. B. C. de Melo, T. Frederico, E. Pace and 
G. Salm\`e}

\newpage 

\begin{figure}
\psfig{bbllx=20mm,bblly=30mm,bburx=100mm,bbury=220mm,file=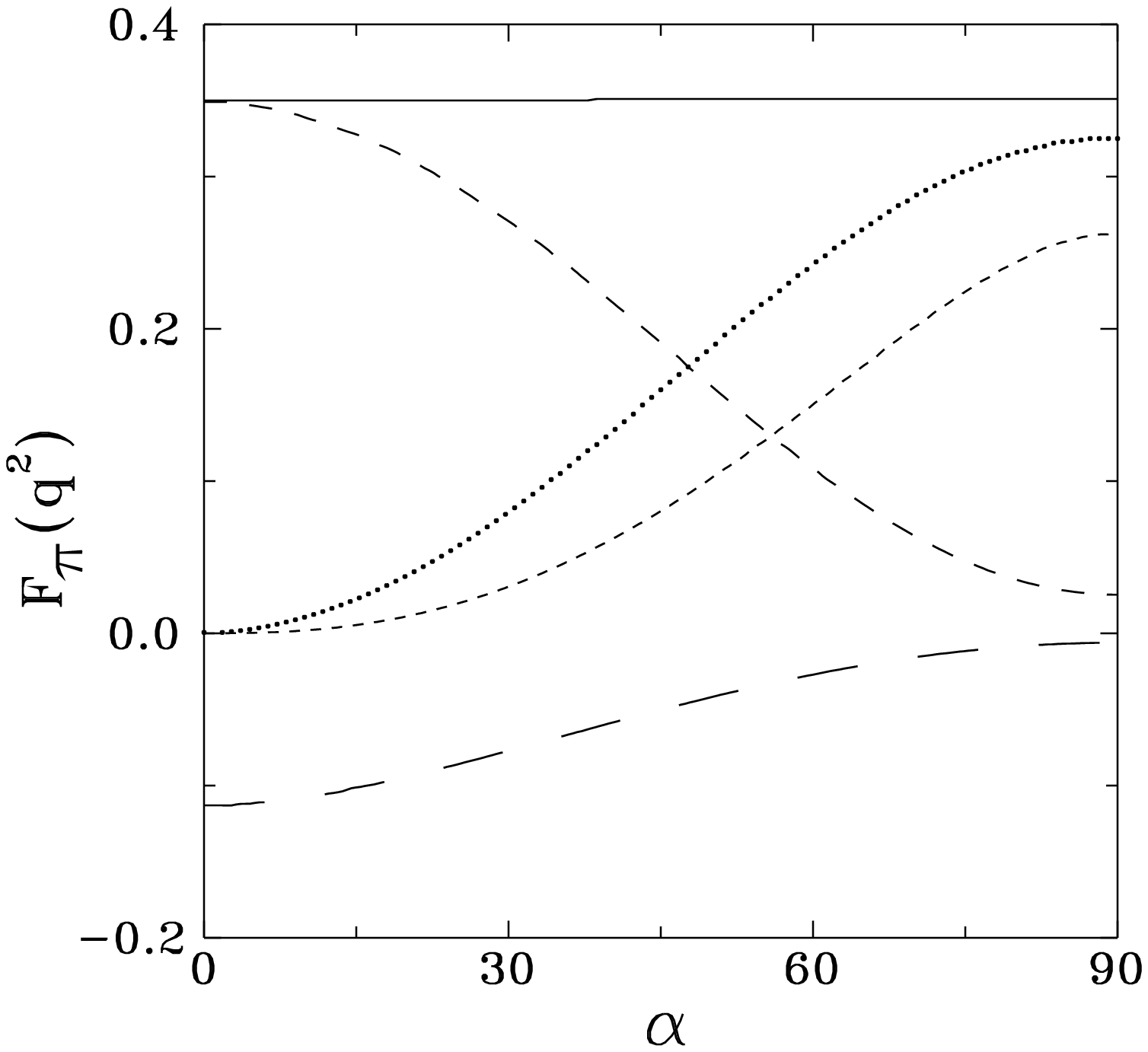}
\end{figure}
\vspace{2cm}

{\bf Fig. 3 - J. P. B. C. de Melo, T. Frederico, E. Pace and 
G. Salm\`e}

\newpage 

\begin{figure}
\psfig{bbllx=20mm,bblly=30mm,bburx=100mm,bbury=220mm,file=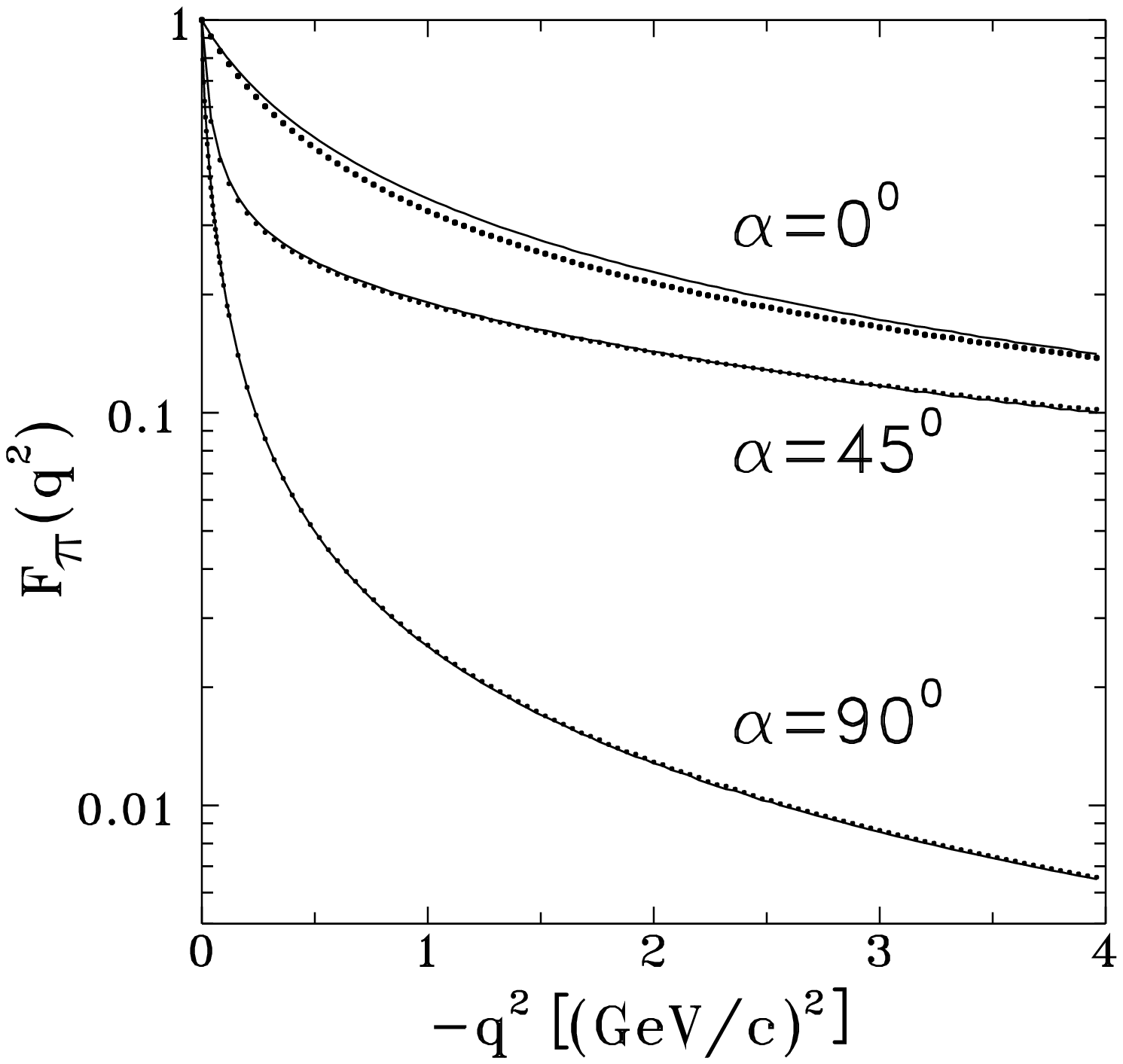}

\end{figure}
\vspace{2cm}

{\bf Fig. 4 - J. P. B. C. de Melo, T. Frederico, E. Pace and 
G. Salm\`e}

\newpage

\begin{figure}
\psfig{bbllx=20mm,bblly=30mm,bburx=100mm,bbury=220mm,file=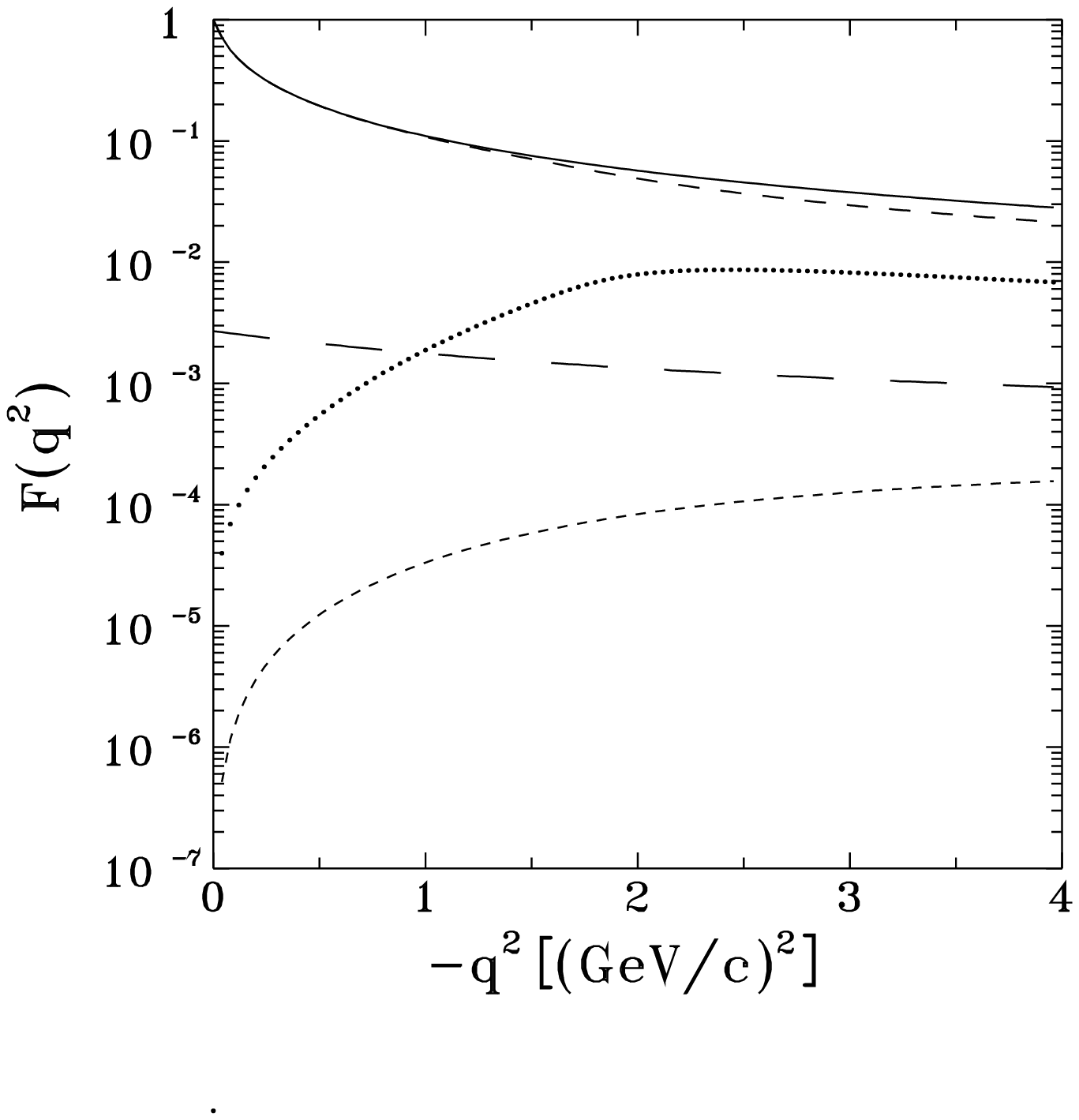}

\end{figure} 
\vspace{2cm}

{\bf Fig. 5 - J. P. B. C. de Melo, T. Frederico, E. Pace and 
G. Salm\`e}

\newpage

\begin{figure}
\psfig{bbllx=20mm,bblly=30mm,bburx=100mm,bbury=220mm,file=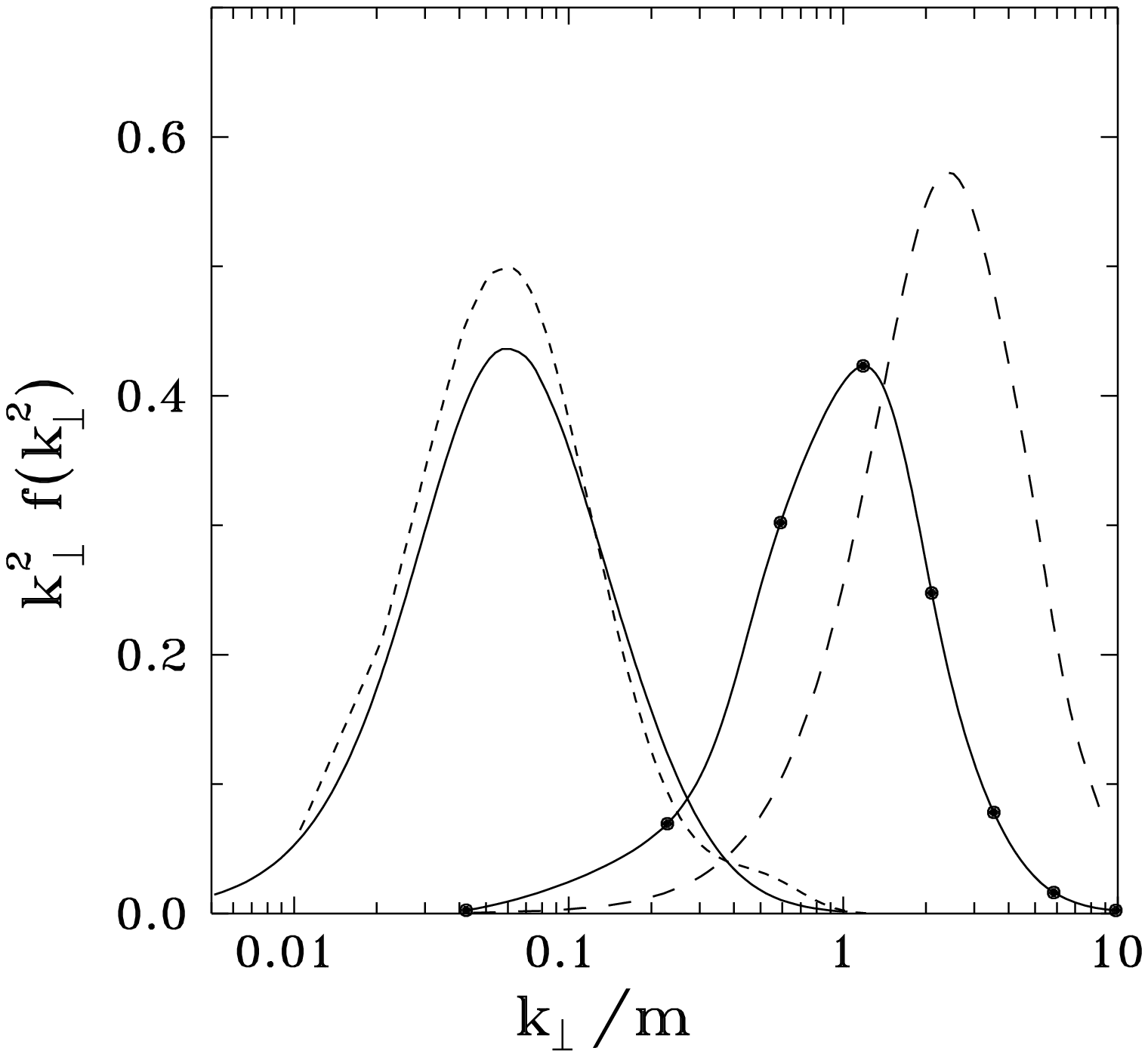}

\end{figure}

\vspace{2cm}

{\bf Fig. 6 - J. P. B. C. de Melo, T. Frederico, E. Pace and 
G. Salm\`e}

\newpage

\begin{figure}
\psfig{bbllx=20mm,bblly=30mm,bburx=100mm,bbury=220mm,file=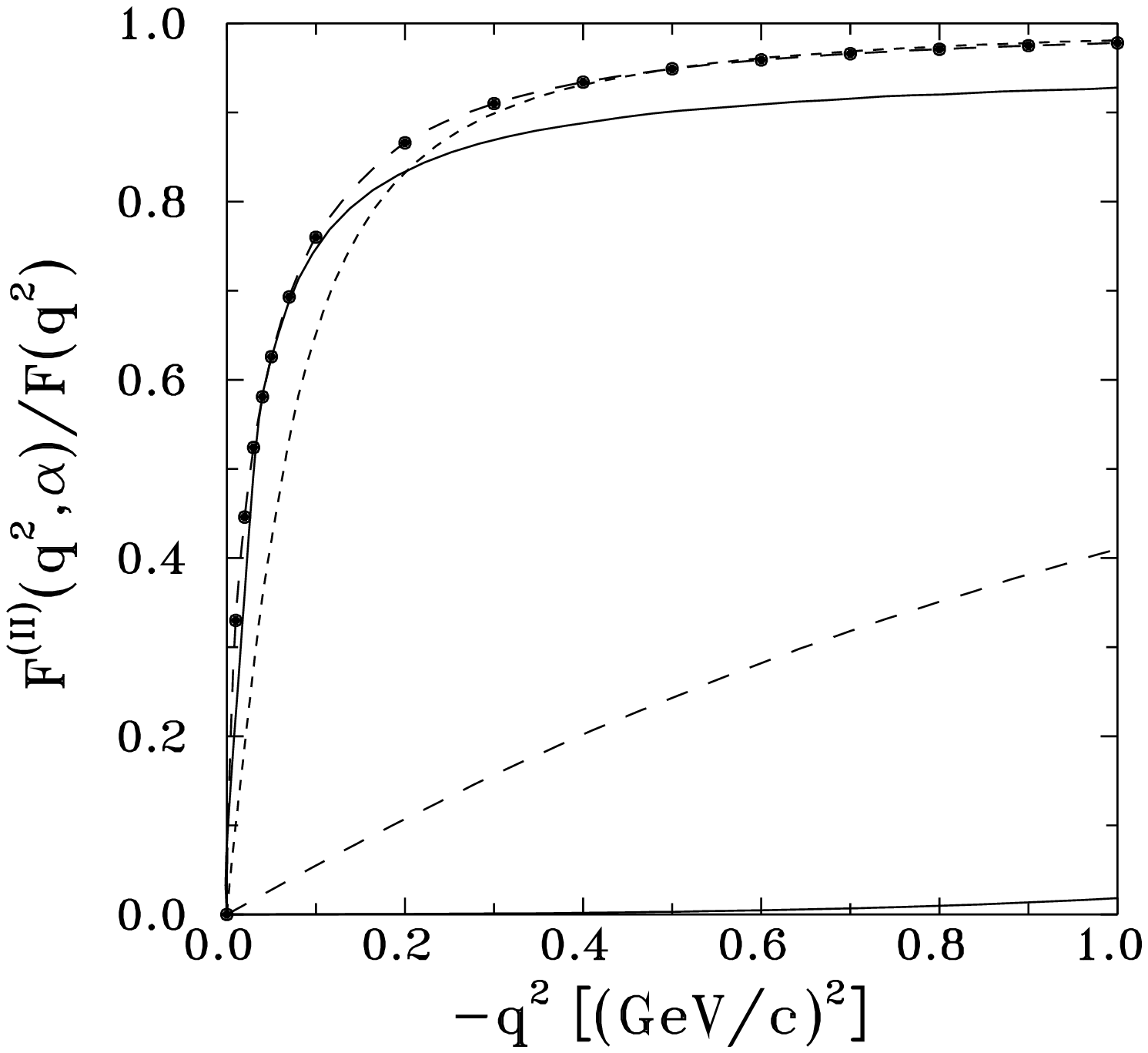}

\end{figure}
\vspace{2cm}

{\bf Fig. 7 - J. P. B. C. de Melo, T. Frederico, E. Pace and 
G. Salm\`e}

\end{document}